\documentclass{article}

\font\tenrm=cmr10

\font\elevenbf=cmbx10 scaled\magstep 1
\font\elevenrm=cmr10 scaled\magstep 1
\font\elevenit=cmti10 scaled\magstep 1

\renewenvironment{thebibliography}[1]
 { \elevenrm
   \begin{list}{\arabic{enumi}.}
    {\usecounter{enumi} \setlength{\parsep}{0pt}
     \setlength{\itemsep}{3pt} \settowidth{\labelwidth}{#1.}
     \sloppy
    }}{\end{list}}

\def\IH{\relax{\rm I\kern-.18em H}}
\font\cmss=cmss10 \font\cmsss=cmss10 at 7pt
\def\ZZ{\relax\ifmmode\mathchoice
{\hbox{\cmss Z\kern-.4em Z}}{\hbox{\cmss Z\kern-.4em Z}}
{\lower.9pt\hbox{\cmsss Z\kern-.4em Z}}
{\lower1.2pt\hbox{\cmsss Z\kern-.4em Z}}\else{\cmss Z\kern-.4em
Z}\fi}
\newcommand{\Dsl}{\not\!\! D}
\newcommand{\dsl}{\not\!\partial}
\newcommand{\bbox}{\lower.2ex\hbox{$\Box$}}
\newcommand{\eqn}[1]{(\ref{#1})}
\newcommand{\ft}[2]{{\textstyle\frac{#1}{#2}}}

\csname @addtoreset\endcsname{equation}{section}
\newsavebox{\uuunit}
\sbox{\uuunit}
    {\setlength{\unitlength}{0.825em}
     \begin{picture}(0.6,0.7)
        \thinlines
        \put(0,0){\line(1,0){0.5}}
        \put(0.15,0){\line(0,1){0.7}}
        \put(0.35,0){\line(0,1){0.8}}
       \multiput(0.3,0.8)(-0.04,-0.02){12}{\rule{0.5pt}{0.5pt}}
     \end {picture}}
\newcommand {\unity}{\mathord{\!\usebox{\uuunit}}}
\newcommand  {\Rbar} {{\mbox{\rm$\mbox{I}\!\mbox{R}$}}}

\newcommand {\Cbar}
    {\mathord{\setlength{\unitlength}{1em}
     \begin{picture}(0.6,0.7)(-0.1,0)
        \put(-0.1,0){\rm C}
        \thicklines
        \put(0.2,0.05){\line(0,1){0.55}}
     \end {picture}}}
\newsavebox{\zzzbar}
\sbox{\zzzbar}
  {\setlength{\unitlength}{0.9em}
  \begin{picture}(0.6,0.7)
  \thinlines
  \put(0,0){\line(1,0){0.6}}
  \put(0,0.75){\line(1,0){0.575}}
  \multiput(0,0)(0.0125,0.025){30}{\rule{0.3pt}{0.3pt}}
  \multiput(0.2,0)(0.0125,0.025){30}{\rule{0.3pt}{0.3pt}}
  \put(0,0.75){\line(0,-1){0.15}}
  \put(0.015,0.75){\line(0,-1){0.1}}
  \put(0.03,0.75){\line(0,-1){0.075}}
  \put(0.045,0.75){\line(0,-1){0.05}}
  \put(0.05,0.75){\line(0,-1){0.025}}
  \put(0.6,0){\line(0,1){0.15}}
  \put(0.585,0){\line(0,1){0.1}}
  \put(0.57,0){\line(0,1){0.075}}
  \put(0.555,0){\line(0,1){0.05}}
  \put(0.55,0){\line(0,1){0.025}}
  \end{picture}}
\newcommand{\Zbar}{\mathord{\!{\usebox{\zzzbar}}}}
\def\IP{\relax{\rm I\kern-.18em P}}
\newcommand{\cmap}{{\elevenbf c} map}
\newcommand{\rmap}{{\elevenbf r} map}
\newcommand{\crmap}{{{\elevenbf c}$\scriptstyle\circ${\elevenbf r} map}}
\newcommand{\Ka}{K\"ahler}
\newcommand{\qu}{quaternionic}

\def\jb{{\bar \jmath}}

\newcommand{\G}{\Gamma}
\newcommand{\del}{\partial}
\def\cF{{\cal F}}

\def\cN{{\cal N}}

\def\IZ{{\hbox{{\rm Z}\kern-.4em\hbox{\rm Z}}}}
\def\Im{{\rm Im ~}}
\def\Re{{\rm Re ~}}
\def\dop{{\rm d}}
\newcommand{\dXN}[1]
{\dop X^{\Sigma_{#1}}\wedge\cdots\wedge\dop X^{\Sigma_N}}

\renewcommand{\section}[1]{\refstepcounter{section}
\vglue 0.5cm
{\elevenbf\noindent \thesection . #1}
\vglue 0.4cm
\setcounter{subsection}{0}
}
\newcommand{\sectionsub}[1]{\refstepcounter{section}
\vglue 0.5cm
{\elevenbf\noindent \thesection . #1}
\vglue 0.2cm
  \setcounter{subsection}{0}}
\renewcommand{\subsection}[1]{\refstepcounter{subsection}
\vglue 0.2cm
{\elevenit \noindent \thesubsection . #1}
\vglue 0.1cm
 }
\begin{document}
\begin{titlepage}
\begin{flushright} KUL-TF-95/39 \\ hep-th/9512139
\end{flushright}
\vfill
\begin{center}
{\LARGE\bf Vector multiplets in $N=2$ supersymmetry\\
and its associated moduli spaces }$^1$   \\
\vskip 27.mm  \large
{\bf   Antoine Van Proeyen}$^2$ \\
\vskip 1cm
{\em Instituut voor theoretische fysica}\\
{\em Universiteit Leuven, B-3001 Leuven, Belgium}
\end{center}
\vfill

\begin{center}
{\elevenbf ABSTRACT}
\end{center}
\begin{quote}
An introduction to $N=2$ rigid and local supersymmetry is given. The
construction of the actions of vector multiplets is reviewed,
defining special K\"ahler manifolds. Symplectic transformations lead to
either isometries or symplectic reparametrizations. Writing down a
symplectic formulation of special geometry clarifies the relation to
the moduli spaces of a Riemann surface or a Calabi-Yau 3-fold. The
scheme for obtaining perturbative and non-perturbative corrections
to a supersymmetry model is explained. The Seiberg-Witten model is
reviewed as an example of the identification of duality symmetries
with monodromies and symmetries of the associated moduli space of a
Riemann surface.
\vfill      \hrule width 5.cm
\vskip 2.mm
{\small
\noindent $^1$ Lectures given in the 1995 Trieste summer school in
high energy physics and cosmology.
\\
\noindent $^2$ Onderzoeksleider NFWO, Belgium; E-mail:
Antoine.VanProeyen@fys.kuleuven.ac.be}
\end{quote}
\begin{flushleft}
November 1995
\end{flushleft}
\end{titlepage}
\begin{center}
{\elevenbf\large VECTOR MULTIPLETS IN $N=2$ SUPERSYMMETRY\\
AND ITS ASSOCIATED MODULI SPACES}\\[15mm]
Antoine VAN PROEYEN    \\[2mm]
{\em Instituut voor theoretische fysica}\\
{\em Universiteit Leuven, B-3001 Leuven, Belgium}\\[1cm]
ABSTRACT
\end{center} \vspace{-3mm}
\begin{quote}
 \baselineskip=12pt \tenrm
An introduction to $N=2$ rigid and local supersymmetry is given. The
construction of the actions of vector multiplets is reviewed,
defining special K\"ahler manifolds. Symplectic transformations lead to
either isometries or symplectic reparametrizations. Writing down a
symplectic formulation of special geometry clarifies the relation to
the moduli spaces of a Riemann surface or a Calabi-Yau 3-fold. The
scheme for obtaining perturbative and non-perturbative corrections
to a supersymmetry model is explained. The Seiberg-Witten model is
reviewed as an example of the identification of duality symmetries
with monodromies and symmetries of the associated moduli space of a
Riemann surface.
\end{quote}
\section{Introduction}
\baselineskip=14pt
\elevenrm

When discussing supersymmetry, it is natural to consider extended
supersymmetries. In the early days of supergravity, one
realized immediately that the maximal extended theory in 4 dimensions
has $N=8$ extended supersymmetry. The idea was therefore that this
should be the ultimate theory. But it turned out that 8 supersymmetries
restrict the theory too much, such that physically
interesting theories could not be constructed. Any
phenomenological theory was based on simple $N=1$. Of course then the
first extension, $N=2$, was considered. It was found that the \Ka\
structure known from $N=1$ appears again, but in a restricted form
\cite{DWLPSVP,DWVP}.

The appearance of geometries, mentioned above, is related to the
spinless fields appearing in these supergravity theories.
They define a map from the
$d$-dimensional Minkowskian space-time to some `target space'
whose metric is given by the kinetic terms of these scalars.
Supersymmetry severely restricts the
possible target-space geometries. The type of target space which one
can obtain depends on $d$ and on $N$, the latter indicating the number of
independent supersymmetry transformations. The number of
supersymmetry generators (`supercharges') is thus equal to $N$ times
the dimension of the (smallest) spinor representation. For realistic
supergravity this number of supercharge components cannot exceed 32.
As 32 is the number of components of a Lorentz spinor in $d=11$
space-time dimensions, it follows that realistic supergravity
theories can only exist for dimensions $d\leq 11$. For the physical
$d=4$ dimensional space-time, one can have supergravity theories
with $1\leq N\leq 8$.

\begin{table}[hbt]
\caption{\tenrm Restrictions on target-space manifolds
according to the type of supergravity theory. The rows are arranged
such that the number $\kappa$ of supercharge components is constant.
${\cal M}$ refers to a general Riemannian
manifold, $SK$ to `special \Ka', $VSR$ to
`very special real' and $Q$ to \qu\ manifolds. }
\label{tbl:mandN}  \begin{center}
\begin{tabular}{c|ccccc}\hline
$\kappa$& $d=2$ & $d=3$ & $d=4$ & $d=5$ & $d=6$  \\ \hline
&$N=1$     &&&& \\
2&${\cal M}$ &&&&  \\   \hline
&$N=2$ &  $N=2$  & $N=1$ &   &  \\
4&\Ka   &  \Ka  & \Ka   &    &    \\ \hline
&$N=4$ &  $N=4$ & $N=2$        & $N=2$    & $N=1$     \\
8&$Q $  &   $Q$   &  $SK\oplus Q$ & $VSR\oplus Q$ & $ Q $   \\
\hline  \hline
&...    & ... & $N=4$  & ... & $\rightarrow $     \\
16&...    & ... & $\frac{SO(6,n)}{SO(6)\otimes SO(n)}\otimes
\frac{SU(1,1)}{U(1)}$  & ... & $d=10$\\ \hline
&...    & ... & $N=8$  & ... &   $\rightarrow $ \\
32&...    & ... & $\frac{E_7}{SU(8)}$  & ... & $d=11$ \\ \hline
\end{tabular} \end{center}
\end{table}

As clearly exhibited in table~\ref{tbl:mandN}, the more supercharge
components one has, the more restrictions one finds. When the number
of supercharge components exceeds 8, the target spaces are
restricted to symmetric spaces. For $\kappa=16$ components, they are
specified by $n$, the number of vector multiplets. This row
continues to $N=1$, $d=10$. Beyond 16 supercharge components there
is no freedom left. The row with 32 supercharge components continues
to $N=1$, $d=11$. Here I treat the case of 8 supercharge components.
This is the highest value of $N$ where the target space is not
restricted to be a symmetric space, although supersymmetry has
already fixed a lot of its structure. This makes this row rather
interesting. For physical theories we are of course mostly interested
in $N=2$ in 4 dimensions. Below I will give some more motivations for
studying $N=2$ in $d=4$. First I review the physical contents of these
theories.
\par
\begin{table}[hbt]
\caption{\tenrm Multiplets of $N=2$, $d=4$}\label{tbl:multN2d4}
\begin{center}
\begin{tabular}{|c|ccc|}\hline
spin& pure SG &$n$ vector m. &$s$ hypermult. \\ \hline
2& 1& & \\
$3/2$& 2& & \\
1&1 &n & \\
$1/2$&&2n & 2s \\
0&&2n& 4s \\ \hline
\end{tabular} \end{center}
\end{table}
The physical fields which occur in $N=2$, $d=4$ are shown in
table~\ref{tbl:multN2d4}.
In rigid supersymmetry there can be vector multiplets and
hypermultiplets. In the former, the vector potentials, which
describe the spin-1 particles, are accompanied by complex scalar
fields and doublets of spinor fields. The vectors can gauge a group,
and all these fields then take values in the associated Lie algebra.
As I will show in section~\ref{ss:rigidN2}, the manifold of the
scalar fields is \Ka ian \cite{PKTN2}.
The hypermultiplets contain a multiple of
4 scalars, and one can construct a \qu\ structure on this space. In
rigid supersymmetry this space is hyper\Ka ian. When coupling to
supergravity an extra spin-1 field, called the graviphoton, appears.
We will see in section~\ref{ss:sugra} what the consequences are of
mixing the vectors in the vector multiplets with the graviphoton. In
supergravity the space of the scalars of hypermultiplets becomes a
\qu\ manifold. The total scalar manifold is factorized into the \qu\
and the \Ka\ manifold. The latter is of a particular type
\cite{DWLPSVP,DWVP}, called {\elevenit special}\footnote{The
terminology `special
\Ka ' is used for this structure which one finds in the context of
supergravity. The manifolds obtained in rigid supersymmetry are
usually called `rigid special \Ka\ manifolds'.} \cite{special}.
Recently the special \Ka\ structure received a lot of attention,
because it plays an important role in string compactifications. Also
\qu\ manifolds appear in this context, and also here it is a
restricted class of special \qu\ manifolds that is relevant. In
lowest order of the string coupling constant these manifolds are
even `very special' \Ka\ and \qu, a notion that will be defined in
section~\ref{ss:sugra}.
\par
A further motivation for $N=2$ theories comes from the
compactification of superstring theories. It turns out that $N=2$
often appears in this context.
 The amount of supersymmetry one gets in compactifications of
 superstrings depend on the type of superstring one starts with, and
 on the compactification manifold, see table~\ref{tbl:stringcomp}.
\begin{table}[ht]\caption{\tenrm String compactification and supersymmetry}
\label{tbl:stringcomp}\begin{center}\begin{tabular}{||c|cc||}\hline
& Heterotic & Type II   \\ \hline
(2,0)& $N=1$ &          \\
(2,1)&       & $N=1$    \\
(2,2)& $N=1$ & $N=2$    \\
(4,0)& $N=2$ & $N=2$    \\
(4,2)& $N=2$ & $N=3$    \\
(4,4)& $N=2$ & $N=4$    \\ \hline
\end{tabular}\end{center}\end{table}
The compactification manifolds are classified here by the number of
left and right handed world-sheet supersymmetries. (Note that
possibly only a part of the compactification manifold has the world-sheet
supersymmetries mentioned in the first column).
Calabi--Yau manifolds have (2,2) supersymmetry, while $K3\times
T^2$ is at least a (4,0) compactification. It is clear from this
table that $N=2$ often appears.

Another motivation for $N=2$ comes from topological theories. These
can be constructed by `twisting $N=2$' theories \cite{Wittentop,%
Vafatop,AnselmiFre}. In the simplest version the orignial $
SO(4)_{spin} = SU(2)_L \, \otimes \, SU(2)_R$ Lorentz group is
deformed to $ SO(4)_{spin}^{\prime} = SU(2)_L \, \otimes \,
SU(2)_R^\prime $ where $SU(2)_R^\prime $ is the diagonal subgroup of
$ SU(2)_I \, \otimes \, SU(2)_R$, and $SU(2)_I $ is the group
rotating the 2 supersymmetry charges. Then one generator of $N=2$
modifies the BRST transformations, and an R--symmetry which is
present in $N=2$ modifies the ghost number such that the cohomology
at zero ghost number of the BRST gives the interesting topological
configurations. In \cite{Leuvtop} the ghost number assignments for
this procedure has been improved by splitting Vafa's BRST operator
\cite{Vafatop} in a BRST and anti-BRST. There are also more
complicated twisting procedures \cite{AnselmiFre} which deform also
the $SU(2)_L$. In any case the correlation functions of $N=2$
theories can be connected to topological quantities. The application
of this procedure in compactified string theories has been discussed
in \cite{rtwist}.

The final motivation I present, has to do with the recent ideas about
dualities. This is the main issue in the lectures below.
I already mentioned that
the restrictions from the presence of 2 supersymmetries still
allow enough freedom for the manifold to be not restricted to a
symmetric space. The definition of these manifolds then depends on
some arbitrary functions, as the \Ka\ potential in $N=1$. We will see
in the coming lectures that the special \Ka\ manifolds are determined
by a holomorphic function, which is called a prepotential. This
restriction was crucial to obtain exact quantum results for $N=2$
theories in the famous Seiberg-Witten papers last year \cite{SeiWit}.

The general idea for obtaining such exact results by connecting the
$N=2$ theory to moduli of surfaces is as follows.
The $N=2$ theories (Seiberg and Witten considered rigid $N=2$
supersymmetry) have a potential that is zero for arbitrary values
of some (massless) scalars. The value of these scalars therefore
parametrize the vacua of these theories. The aim is then to find an
effective quantum theory for these scalars when the massive states
are integrated out of the path integral. This effective action
gets perturbative and non-perturbative contributions. It will be
invariant under a set of duality symmetries, which are a subset of
symplectic transformations as I will discuss in
section~\ref{ss:sympltr}. After getting clues from the perturbative
results, Seiberg and Witten have connected the scalar fields to
moduli of a Riemann surface, and conjecture that the full quantum
theory can be obtained from a metric in the space of moduli. The
duality transformations of the effective quantum theory are now
represented either as symmetries of the defining equation of the
Riemann surface or as transformations obtained by encircling singular
points in moduli space, called monodromies. These singular points
correspond in the effective field theory to vacua where other states
become massless, and therefore at these points the starting setup was
not valid, leading to the singularity.

In supergravity theories the same ideas can be applied. The main
difference is now that the connection should be made to moduli of
Calabi-Yau 3-folds \cite{modssym}. Moreover, in this case the
surface is not just a geometrical tool, but can be seen as target
space of the dual theory in the context of string compactifications.

For that purpose consider compactifications of the heterotic string
on $K3\times T^2$ manifolds on the one hand, and type II strings
(type IIA for definiteness) compactified on a Calabi-Yau manifold on
the other hand. Both lead to $N=2$ theories with vector and
hypermultiplets. For the Calabi-Yau manifold in type IIA one obtains
$h^{1,1}$ neutral vectormultiplets and $h^{1,2}+1$ neutral
hypermultiplets. (The Hodge numbers are interchanged for type IIB).
Now there are important facts of $N=2$ which come
to help to get information on the quantum theories.
First, the scalar `dilaton' field of the original superstring action
becomes part of a vector multiplet when compactifying a heterotic
string, while it becomes part of a hypermultiplet when compactifying
a type II string. This can already be seen from counting the number
of fields in the compactified theory.
Secondly, this dilaton field $S$ arranges the
perturbation theory. Its expectation value is
$\langle S\rangle =\frac{\theta}{2\pi}+\frac{i}{g^2}$, where $\theta$
is the theta-angle and $g$ is the coupling constant, which should
appear in all quantum corrections. The third information is that in
$N=2$ there are no couplings between the scalars of vector multiplets
and those of neutral hypermultiplets. Combining these 3 facts, one
concludes that in the compactified heterotic string the
hypermultiplet manifold is not quantum corrected as it can not depend
on the coupling constant. For the compactified type II string this
holds for the vector multiplets.
The further assumption is the validity of the duality hypothesis, called
`second quantised mirror symmetry' \cite{2ndqms}. This states that the
quantum theories of the mentioned heterotic and type II compactified
theories are dual to each other. This hypothesis is used to get
information about the vector
multiplet couplings which can be obtained through dimensional
reduction of the heterotic string. These are thus related to the
quantum theory of the type II theory, which is by the previous
arguments the same as the classical theory. That classical theory is
the one of the moduli of Calabi-Yau manifolds. Therefore we have here
a string-based relation between the quantum theory of vector
multiplet couplings and the moduli space of Calabi-Yau manifolds.
\vspace{1mm}
\par
Section~\ref{ss:rigidN2} will treat rigid $N=2$ supersymmetry.
I will pay most attention to the vector multiplets,
explaining their description in superspace. Their
action is determined by a holomorphic prepotential.
Duality symmetries (symplectic transformations) are first shown for
general couplings of scalars and vectors in section~\ref{ss:sympltr}.
Then I specify
to the case of $N=2$. There are two kind of applications, either as
isometries of the manifolds (symmetries of the theory), or as
equivalence relations of prepotentials (pseudo-symmetries). As an
example, I will look at the duality symmetries of the Seiberg-Witten
model when the perturbative quantum corrections
are taken into account. In section~\ref{ss:symplrigidsg}
I will give
another definition of the geometry of the scalar manifold in a
coordinate independent way, and covariant for symplectic transformations,
which paves the way for the comparison with the theory of the
space of moduli of a Riemann surface (section~\ref{ss:RS}). This
moduli space is
conjectured to describe the full quantum theory for the massless fields.
Section~\ref{ss:solSW} will contain this theory for the
simplest example of Seiberg and Witten \cite{SeiWit}.
In section~\ref{ss:sugra} I will exhibit how the
structure gets more rich in supergravity, where the space is
embedded in a projective space. This structure was found by starting
from the superconformal tensor calculus. The symplectic formulation
and connection to Calabi-Yau moduli will be explained shortly, mainly
by making the analogy with the rigid case. Finally also the basic
facts about special quaternionic and
classification of homogeneous special manifolds are recalled.

In a first appendix the conventions are explained and some useful
formulae are given. The second appendix contains a translation table
for conventions used in the $N=2$ literature. The final two appendices
are related to technical issues for section~\ref{ss:RS}: the volume
form in pseudo homogeneous spaces and some main facts of elliptic
integrals.
\pagebreak[2]
\sectionsub{Rigid $N=2$} \label{ss:rigidN2}
\subsection{The $N=2$ algebra}
Supersymmetry by definition means that the supersymmetry operator
$Q_\alpha$ squares to the momentum operator $P_\mu$. For several
supersymmetries, labelled by $i$, the algebra is
\begin{equation}
\{ Q_\alpha^i, Q_{\beta j}\}=-i\left( \gamma^\mu {\cal C}^{-1}\right)
_{\alpha\beta}P_\mu \delta^i_j\ . \label{susyalgebra}
\end{equation}
The factor $i$ is introduced for consistency with hermitian
conjugation with an hermitian $P_\mu$. See the appendix for hermitian
conjugation of the spinors, where it is also explained that the position
of the index $i$ indicates the chirality. The hermitian conjugates of
the generators
\begin{equation}
Q^{\dagger \alpha i}\equiv
\left( Q_{\alpha i}\right)^\dagger = -i Q^{i\,T} {\cal C}\gamma_0
\label{Qdagger}\end{equation}
then satisfy
\begin{equation}
\left\{ Q_{\alpha i}, Q^{\dagger \beta j}\right\}= P_\mu \delta^j_i
\left( \gamma^\mu \gamma_0\right) _\alpha{}^\beta \ .
\end{equation}
With $P_0$ the energy, or the mass, this exhibits the positive
energy statements in supersymmetry. To see this, select two values of
$\alpha$
for which $Q_\alpha$ are considered as annihilation operators, and
$Q^{\dagger \alpha}$ are then creation operators (they are related by
\eqn{Qdagger} to the other two $Q_\alpha$).

The Haag-Lopuszanski-Sohnius theorem \cite{HLS} restricts the
symmetries which a field theory  with non-trivial scattering
amplitudes can have. It restricts then also the algebra of
the supersymmetries, but still allows central charges in extended
supersymmetry. This means that we may still have (for $N=2$)
\begin{equation}
\left\{ Q_{\alpha i}, Q_{\beta j}\right\}= {\cal
C}^{-1} _{\alpha\beta}\epsilon_{ij} Z\ .   \label{Zalgebra}
\end{equation}
For higher extended supersymmetry one could have $Z_{ij}$
antisymmetric, which for $N=2$ reduces to the mentioned form. For the
hermitian conjugates this implies
\begin{equation}
\left\{ Q^{\dagger \alpha i}, Q^{\dagger \beta j}\right\}=-\epsilon^{ij}
 {\cal C} ^{\alpha\beta} Z^\dagger \ .
\end{equation}
Remark that this theorem concerns algebras of symmetries
in the sense that the structure constants are constant, and not
field-dependent 'structure functions' as in the 'soft algebras'
that are used in field representations of supersymmetry. The
theorem should then apply for the vacuum expectation values of the
structure functions.

Define now
\begin{equation}
A_{\alpha i}\equiv Q_{\alpha i}+e^{i\theta}
\epsilon_{ij}{\cal C}^{-1}_{\alpha\beta}
Q^{\dagger \beta j}\ ,
\end{equation}
where $e^{i\theta}$ is an arbitrary phase factor.
 The hermitian conjugates
are (${\cal C}$ is taken to be unitary)
\begin{equation}
A^{\dagger \alpha i}=Q^{\dagger \alpha i}+
e^{-i\theta}\epsilon^{ij}Q_{j\beta}
{\cal C}^{\beta\alpha} =e^{-i\theta} \epsilon^{ij}A_{j\beta}
{\cal C}^{\beta\alpha}\ .
\end{equation}
Consider now their anticommutators on a state for which
$P_\mu =\delta_\mu^0 M$:
\begin{equation}
\left\{ A_{\alpha i},A_{\beta j}\right\}=
\epsilon_{ij}{\cal C}^{-1}_{\alpha\beta} (Z +2e^{i\theta} M+Z^\dagger
e^{2i\theta})
\end{equation}
 or \begin{equation}
\left\{ A_{\alpha i},A^{\dagger \beta j}\right\}= \delta_i^j
\delta_\alpha^\beta (2M+ Ze^{-i\theta}+Z^\dagger e^{i\theta})\ .
\end{equation}
As the left hand side is a positive definite operator, the
right hand side should be positive for all $\theta$, which
shows that $M\geq |Z|$. This is an important result relating the
masses to the central charges. One can now also show that if the
equality is satisfied for a state, then that state is invariant
under some supersymmetry operation.

I will not require the presence of central charges,
but we will find that they are needed for certain representations.
\subsection{Multiplets}    \label{ss:multiplets}
The superfield which is most useful for our purposes is the chiral
superfield. The $N=2$ superspace is built with anticommuting
coordinates $\theta^i_\alpha$, and $\theta_{\alpha i}$, where again
the position of the index $i$ indicates the chirality. A chiral
superfield is defined by a constraint $ D^{ \alpha i}\Phi=0 $, where
$D^{\alpha i}$ indicates a (chiral) covariant derivative in
superspace. I will not give a detailed definition of covariant
derivatives in superspace as this will not be necessary for the
following. The superfield $\Phi$ is complex, and can be
expanded as
\[ \Phi= A + \theta^i_\alpha  \Psi_i^\alpha +
{\cal C}^{\alpha\beta} \theta^i_\alpha  \theta^j_\beta B_{ij}+
\epsilon_{ij} \theta^i_\alpha  \theta^j_\beta  F^{\alpha\beta}
+\ldots \ .\]
$B_{ij}$ is symmetric and so is $F^{\alpha\beta}$, which can then,
due to the chirality and the symmetry, be written as
$F^{\alpha\beta}= \sigma^{\alpha\beta} _{ab}F^{ab\, -}$, where
$F^{ab\,-}$ is an arbitrary antisymmetric antiselfdual tensor
(the selfdual part occurs in $\bar \Phi$).

One can define a chiral multiplet also without superspace. Then
one starts from a complex scalar $A$ and demands that it transforms
under supersymmetry only with a chiral supersymmetry parameter (no
terms with $\epsilon_i$):
\begin{equation}
\delta(\epsilon) A= \bar \epsilon^i \Psi_i \ . \label{delA}
\end{equation}
The spinor $\Psi_i$ is a new field, and one then takes for $\Psi$ the
most general transformation law compatible with the supersymmetry
algebra. First, I translate the anticommutator of the generators
$Q_{\alpha i}$, \eqn{susyalgebra}, to a commutator of
$\delta(\epsilon)\equiv\bar \epsilon^i Q_i +\bar
\epsilon_i Q^i$. The operator $P_\mu$ is represented on fields as
$P_\mu=-i\partial_\mu$. This leads to
\begin{equation}
\left[ \delta(\epsilon_1),\delta(\epsilon_2)\right]=\bar \epsilon_2^i
\gamma^\mu \epsilon_{1i}\partial_\mu + \bar \epsilon_{2i}
\gamma^\mu \epsilon_1^i\partial_\mu \ ,\label{algepsilon}
\end{equation}
where the second term is the hermitian conjugate of the first, or
could also be denoted as $-(1\leftrightarrow 2)$.

Compatibility with this algebra restricts $\delta\Psi$ to
\begin{equation}
\delta(\epsilon)\Psi_i=\dsl A \epsilon_i +
\ft12 B_{ij}\epsilon^j+\ft12 \sigma_{ab}F^{-ab}\epsilon_{ij}\epsilon^j\ ,
\label{deltaPsi}\end{equation}
where $B_{ij}$ is an arbitrary symmetric field as in the superspace,
and $F^{ab}$ is
antisymmetric (and the (anti)selfduality is automatic, see
\eqn{antisdsigma}). This most general transformation law defines the
next components of the multiplets. If we continue with the most
general transformation laws for $B$ and $F$ we find again new
arbitrary fields, and finally the procedure ends defining the chiral
multiplet with as free components
\begin{equation}
\left( A,\Psi_i, B_{ij}, F_{ab} ,\Lambda_i, C\right) \ .
\label{compchiral}
\end{equation}
$\Lambda_i$ is a spinor and $C$ a complex scalar. Note that we did
not allow central charges in the algebra. Allowing these would
already change \eqn{deltaPsi}, see \cite{structure}, and lead to more
components.

In $N=2$ the minimal multiplets have 8+8 real components. The chiral
multiplet has 16+16 components and is a reducible multiplet. The
vector multiplet \cite{vectorm} is an irreducible 8+8 part of this chiral
multiplet. The others form a `linear multiplet', see below. The
reduction is accomplished by an additional constraint, which in
superspace reads
\begin{equation}
D_{\alpha(i} D_{j)\beta} \Phi\,{\cal C}^{\alpha\beta}
= \epsilon_{ik}\epsilon_{j\ell }
 D_\alpha^{ (k}  D_\beta^{\ell ) }\bar  \Phi \,  {\cal C}^{\alpha\beta}
\ ,\label{constrvector}\end{equation}
where $\bar \Phi$ is the complex conjugate superfield, containing the
complex conjugate fields.
In components, this is equivalent to the condition
\begin{equation}
L_{ij}\equiv B_{ij}- \epsilon_{ik}\epsilon_{j\ell }\bar B^{k\ell } =0\ ,
\label{Breal}\end{equation}
and the equations which follow from this by supersymmetry.
These are
\begin{eqnarray}
\phi_i&\equiv &\dsl \Psi_i-\epsilon^{ij}\Lambda_j=0\nonumber\\
E^a&\equiv &\partial_b \,{}^\star F^{ab}=0  \nonumber\\
H&\equiv& C-2\partial_a\partial^a \bar A =0\ . \label{vectorconstr}
\end{eqnarray}
The equation \eqn{Breal} is a reality condition. It leaves in
$B_{ij}$ only 3 free real components. The other equations
define $\Lambda$ and $C$ in terms of $\Psi$ and $A$, and the remaining
one is a Bianchi identity for $F_{ab}$, which implies that this is
the field strength of a
vector potential. The constrained multiplet is
therefore called a `{\elevenit vector multiplet}', which is thus a
multiplet
consisting of independent fields (we give new names
for the independent fields of a vector multiplet)
$\left( X; \Omega_i;Y_{ij};W_\mu\right)$ with transformation laws
\begin{eqnarray}
\delta(\epsilon) X&=&\bar \epsilon^i\Omega_i\nonumber\\
\delta(\epsilon)\Omega_i&=&\dsl X \epsilon_i +
\ft12 Y_{ij}\epsilon^j+\ft12 \sigma_{ab}{\cal F}^{-ab}\epsilon_{ij}
\epsilon^j\nonumber\\
\delta(\epsilon) Y_{ij}&=&2\bar \epsilon_{(i}\dsl\Omega_{j)}+2\epsilon_{ik}
\epsilon_{j\ell }\bar \epsilon^{(k}\dsl \Omega^{\ell )}\nonumber\\
\delta(\epsilon) W_\mu&=&\bar
\epsilon_i\gamma_\mu\Omega_j\epsilon^{ij}+
\epsilon^i\gamma_\mu\Omega^j\epsilon_{ij} \ .
\end{eqnarray}
Here, the vector is abelian, and ${\cal F}_{\mu\nu}=2\partial_{[\mu}
W_{\nu]}$. Let me remark that these $N=2$
multiplets exist in dimensions up to $d=6$. In 6 dimensions, the vector
multiplet has no scalars\footnote{Therefore you find no trace of these
in table~\ref{tbl:mandN}}, and consists of $(\Omega_i;Y_{ij};W_\mu )$,
where now $\mu$ runs over 6 values. We may
understand the complex scalar $X$ in four dimensions as the fifth and
sixth coordinate of the vector in 6 dimensions.

Now, I extend the vector multiplet to the
non--abelian case. This means that the vectors gauge a group $G$ with
generators $T_A$ and
coupling constant $g$. The index $A$ runs over $n\equiv dim\ G$
values. I then attach an index $A$ to all fields of the vector
multiplets, and write the fields in the adjoint as e.g.
$X=X^A T_A$, and (with parameters $y^A$)
\begin{equation}
\delta_G(y) X= g[y,X]=g\,y^A \, X^B\, [T_A,T_B]\ ;\qquad
\delta_G(y)W_\mu= \partial_\mu y +g[y,W_\mu]\  .
\end{equation}
First we could try to replace all derivatives
with derivatives covariant under the gauge group, e.g.
$D_\mu =\partial_\mu-g \delta_G(W_\mu)$. However it can not
be that simple.  Indeed, replacing all derivatives by covariant
derivatives gives also a covariant derivative in \eqn{algepsilon}.
But then the Jacobi identities are not satisfied any more. Indeed,
applying a third supersymmetry on that commutator, this one can also
act on the $W_\mu$ contained in the covariant derivative $D_\mu$.
That gives a term with $\Omega$:
\begin{equation}
\left[\delta(\epsilon_3),\left[\delta(\epsilon_1),\delta(\epsilon_2)
\right]\right] =-  (\bar \epsilon_2^i \gamma^\mu\epsilon_{1i}+h.c.)\
\delta_G( \epsilon_{3k}\gamma_\mu\Omega_j\epsilon^{kj}+h.c.)\ ,
\end{equation}
which violates the Jacobi identity
in $d=4$. In 6 dimensions the Jacobi identity is satisfied, due to a
well-known Fierz identity. Therefore in 6 dimensions, the substitution of
ordinary derivatives by covariant derivatives is sufficient. Reducing
that to 4 dimensions creates terms with the $X$ field. Therefore the
algebra in 4 dimensions is not just \eqn{algepsilon} with a covariant
derivative, but contains an $X$--dependent gauge transformation:
\begin{equation}
\left[ \delta(\epsilon_1),\delta(\epsilon_2)\right]=\bar \epsilon_2^i
\gamma^\mu \epsilon_{1i}D_\mu -2\delta_G\left(
X\bar \epsilon_{1i}\epsilon_{2j} \epsilon^{ij}\right)+h.c.  \ .
\label{naalgepsilon}
\end{equation}
The algebra has become `soft', i.e. the structure functions depend on
fields. If the fields have zero expectation value, then the algebra
reduces in the vacuum to \eqn{susyalgebra}. However,
if some $X^A$ get a non--vanishing vacuum expectation value, the
right hand side of \eqn{Zalgebra} is non--zero, and there is a
central charge $Z=-2\bar \langle X^A\rangle T_A$.

To realise the
new algebra, the transformation rules of the fields have to be modified
by $g$--dependent terms. E.g. for the spinor of the vector multiplet
we now find
\begin{equation}
\delta(\epsilon)\Omega_i = \Dsl X \epsilon_i +\ft12 Y_{ij}\epsilon^j
+\ft12\sigma^{ab}F_{ab}^- \epsilon_{ij}\epsilon^j
-g[X,\bar X] \epsilon_{ij}\epsilon^j \ .
\end{equation}

For a more complete treatment of $N=2$ we should also consider other
multiplets. The physical hypermultiplet can be described by scalar
multiplets \cite{scalarm} consisting of $2s$
complex scalar fields, their central charge transformed fields, which
are auxiliary, and $s$ fermions. An alternative is the linear
multiplet \cite{linearm}. The constraints \eqn{Breal} and
\eqn{vectorconstr} transform as a linear multiplet.
$E_a$ is then the field strength
of an antisymmetric tensor. There is also a non--linear
multiplet \cite{nonlinearm,structure}, a `relaxed hypermultiplet'
\cite{relaxedhyperm}, or descriptions in harmonic
superspace \cite{harmonic}. Similarly there is an
alternative for the vector multiplet where one of the scalars is
described by an antisymmetric tensor \cite{tenvecm,dWKLL}, or another
description of massive vectormultiplets \cite{KarpaczI}. In these
lectures I will restrict myself to the vector multiplet as
described above.
\pagebreak[2]
\subsection{Action for the vector multiplet}
In superspace, actions can be obtained as integrals over the full or
chiral superspace.
So we can get an action from integrating over a chiral superfield.

Remembering that vector multiplets are chiral superfields, we can
form new chiral superfields by taking an arbitrary holomorphic
function of the vector multiplets. If the superfield defining the
vector multiplet is $\Phi$ (which thus satisfies also
\eqn{constrvector}), then we take the integral (and add its complex
conjugate)
\begin{equation}
\int d^4x\int d^4\theta\ F(\Phi)\ + h.c.\ ,  \label{superspaceF}
\end{equation}
for an arbitrary holomorphic function $F$.

The superfield $F(\Phi)$ has by definition as lowest component
$A=F(X)$. The further components are then defined by the
transformation law, which gives, comparing with \eqn{delA}
$\Psi_i= F_A \Omega^A_i$, where I use here and below the
notations
\begin{eqnarray}
F_A(X) &=&\frac{\partial}{\partial X^A}
F(X)\ ;\qquad \bar F_A(\bar X)\,=\,\frac{\partial}
{\partial \bar X^A}\bar F(\bar X) \nonumber\\
F_{AB}&=& \frac{\partial}{\partial X^A} \frac{\partial}{\partial X^B}
F(X)\qquad \ldots\ .
\end{eqnarray}
Calculating the transformation of $\Psi_i$ one finds $B_{ij}$
and $F_{ab}$, \ldots~.
In components, the integral \eqn{superspaceF} is the
highest component of  the superfield, $C$ in the notation of
\eqn{compchiral}.  This leads to
(The standard convention these days is to start with
$A=-iF(X)$, which I now also use):
\begin{eqnarray}
{\cal L}_F&=& -iF_A D_aD^a \bar X^A +\ft i4 F_{AB}{\cal F}_{ab}^{-A}
{\cal F}^{-B\,ab}+ i F_{AB}\bar \Omega_i^A\Dsl
\Omega^{iB}\nonumber\\
&&- \ft i8 F_{AB} Y_{ij}^A Y^{ij\,B}
+\ft i4 F_{ABC} Y^{ij\, A}\bar \Omega_i^B\Omega_j^C\nonumber\\
&&-\ft i4 F_{ABC} \epsilon^{ij}\bar \Omega_i^A \sigma\cdot {\cal F}^{-B}
\Omega_j^C+\ft i{12}F_{ABCD}\epsilon^{ij}\epsilon^{k\ell } \bar \Omega_i^A
\Omega_\ell ^B   \bar \Omega_j^C\Omega_k^D\nonumber\\
&&-ig\epsilon^{ij}F_{AB}\bar \Omega_i^A [\Omega_j,\bar X]^B
-ig F_A [\bar \Omega^i,\Omega^j]^A\epsilon_{ij}
-ig^2F_A\left[\left[\bar X,X\right] ,\bar X\right]^A \nonumber\\
&&+h.c.\label{Lvectorrigid}
\end{eqnarray}
The commutators are meant only for the gauge group, e.g.
\begin{equation}
[\bar \Omega^i,\Omega^j]^A  =f^A_{BC}\bar \Omega^{iB}\Omega^{jC}
\qquad\mbox{with}\qquad [T_B,T_C]=f^A_{BC}T_A\ .
\end{equation}
For the gauging I assumed here that $F$ is an invariant function,
i.e. $\delta F=gF_A [y, X]^A=0$, although a more general situation
exists \cite{dWLVP,toptermsdWHR}. The first terms of the action
give kinetic terms for the scalars $X$, the vectors, and the
fermions $\Omega$. The following term says that $Y_{ij}$ is an
auxiliary field that can be eliminated by its field equation.

Most of the discussion about the supersymmetry actions is only about
their bosonic part. The presence of the fermions and supersymmetry has
given a restriction of the bosonic action, but can then be forgotten
for many considerations. So let me describe what we have obtained
here for the bosonic part.

First, I write a general formulation for the bosonic sector of a
theory with scalar
fields $z^\alpha$, and vector fields labelled
by an index $\Lambda$. If there are no Chern-Simons terms (these do
occur for non--abelian theories if $\delta F\neq 0$), one
can write a general expression
\begin{eqnarray}
{\cal L}_0 &=& -g_{\alpha \bar \beta}D_\mu z^\alpha
D^\mu \bar z^\beta\ -\ V(z) \label{genL01}\\
{\cal L}_1&=&
\ft 12 \Im \left( {\cal N}_{\Lambda\Sigma}(z)\,{\cal F}_{\mu\nu}^{+\Lambda}
 {\cal F}^{+\mu\nu\,\Sigma }\right) =
\ft14(\Im {\cal N}_{\Lambda\Sigma}){\cal F}_{\mu\nu}^\Lambda
{\cal F}^{\mu\nu\Sigma}
-\ft i8 (\Re {\cal N}_{\Lambda\Sigma})
\epsilon^{\mu\nu\rho\sigma}{\cal F}_{\mu\nu}^\Lambda
{\cal F}_{\rho\sigma}^\Sigma \ .   \nonumber
\end{eqnarray}
$g_{\alpha\bar \beta}$ is the (positive definite) metric of the
target space, while $\Im {\cal N}$ is a (negative definite) matrix
of the scalar fields, whose vacuum expectation value gives the gauge
coupling constants, while that of $\Re {\cal N}$ gives the so-called
theta angles. $V$ is the potential.

In our case the $z^\alpha$ can be chosen to be $X^A$ (special
coordinates), and the index $\Lambda$ is also $A$. We obtain
\begin{eqnarray}
G _{A \bar B}(X,\bar X) & = &2\,\Im F_{AB}=
\partial_A\partial_{\bar B}K(X,\bar X)  \qquad\mbox{with }\
K(X,\bar X)=i (\bar F_A(\bar X) X^A-F_A(X) \bar X^A) \nonumber\\
{\cal N}_{AB}&=&\bar F_{AB}
\ ;\qquad V=ig^2F_A\left[\left[\bar X,X\right] ,\bar X\right]^A +h.c.
\label{KFX}
\end{eqnarray}
The metric in target space is thus K\"ahlerian.
For $N=1$ the \Ka\ potential could have been arbitrary. We find here
that the
presence of two independent supersymmetries
implies that this \Ka\ metric, and even the complete action,
depends on
a holomorphic prepotential $F(X)$.
Two different functions $F(X)$ may correspond to equivalent
equations of motion and to the same geometry. It is easy to see that
\begin{equation}
F\approx F +a+ q_A X^A + c_{AB}X^A X^B \ ,\label{FapprF}
\end{equation}
where $a$ and $q_A$ are complex numbers, and $c_{AB}$ real.
But more relations can be derived from the symplectic
transformations that we discuss in section~\ref{ss:sympltr}.

As a simple example, used also in \cite{SeiWit}, consider the $N=2$
susy-YM theory with gauge group $SU(2)$. For invariant holomorphic
function, we can use
\begin{equation}
F=\alpha X^A X^A \ , \label{FSU2}
\end{equation}
where now $A=1,2,3$, and $\alpha$ is a complex number.
As, according to \eqn{FapprF} the real
part of $\alpha$ does not contribute to the action, we can take
$\alpha=i$ (positive imaginary part for positive kinetic energies).
The potential of this theory is
\begin{equation}
V=4g^2\left| \epsilon_{ABC}\bar X^B X^C\right|^2\ .\label{VSU2}
\end{equation}
This shows that it remains zero in valleys where e.g.
$X^A=a\delta^A_3$. Note that different values of $a$ give
different masses for the vectors, so the value of this 'modulus' is
physically relevant. For $a=0$ all the vectors are massless.

The description of the action as it follows from superspace is not
manifestly invariant under reparametrizations of the target space.
Indeed, the superfield constraint \eqn{constrvector} allows only real
linear transformations of the superfields, and thus of the $X^A$. A
description of the scalar manifold covariant under target space
reparametrizations will be given in section~\ref{ss:covrigidsg}. Its
formulation is inspired by the symplectic transformations which we
will find in section~\ref{ss:sympltr}.
\pagebreak[2]
\sectionsub{Symplectic transformations} \label{ss:sympltr}
The symplectic transformations are a generalization of the
electro-magnetic duality transformations. I first recall the
general formalism for arbitrary actions with coupled spin-0 and
spin-1 fields \cite{dual}, and then come to the specific case of $N=2$.
\subsection{Duality symmetries for the vectors}\label{ss:dualvector}
Consider a general action of the form ${\cal L}_1$ in \eqn{genL01}
for abelian spin-1
fields. The field equations for the vectors are
\begin{equation}
0=\frac{\partial{\cal L}}{\partial W_\mu^\Lambda}=2\partial_\nu
\frac{\partial{\cal L}}{\partial {\cal F}^\Lambda_{\mu\nu}}= 2\partial_\nu
\left( \frac{\partial{\cal L}}{\partial {\cal F}^{+\Lambda}_{\mu\nu}}
+ \frac{\partial{\cal L}}{\partial {\cal F}^{-\Lambda}_{\mu\nu}}\right)
\end{equation}
I define
\begin{equation}
  G_{+\Lambda }^{\mu\nu}\equiv 2i\frac{\partial{\cal L}}
  {\partial \cF^{+\Lambda }_{\mu\nu}}=
{\cal N}_{\Lambda \Sigma }\cF^{+\Sigma \,\mu\nu}
\ ;\qquad
  G_{-\Lambda }^{\mu\nu}\equiv -2i\frac{\partial{\cal L}}
  {\partial \cF^{-\Lambda }_{\mu\nu}}=
\bar{\cal N}_{\Lambda \Sigma }\cF^{-\Sigma \,\mu\nu}\ .\label{defG}
\end{equation}
Observe that these relations are only consistent for symmetric
${\cal N}$. So far, this is an obvious remark, as in \eqn{genL01}
we can choose ${\cal N}$ to be symmetric.
The equations for the field strengths can then be written as
\begin{eqnarray}
\del^\mu \Im \cF^{+\Lambda }_{\mu\nu} &=&0\ \ \ \ \ {\rm Bianchi\
identities}\nonumber\\
\del_\mu \Im G_{+\Lambda }^{\mu\nu} &=&0\ \ \ \ \  {\rm Equations\  of\
motion}
\end{eqnarray}
This set of equations is invariant under  $GL(2m,\Rbar)$
transformations:
\begin{equation}
\pmatrix{\widetilde\cF^+\cr \widetilde G_+\cr}={\cal S}
\pmatrix{\cF^+\cr G_+\cr} =
\pmatrix{A&B\cr C&D\cr}   \pmatrix{\cF^+\cr G_+\cr}\ . \label{FGsympl}
\end{equation}
However, the $G_{\mu\nu}$ are related to the ${\cal F}_{\mu\nu}$
as in \eqn{defG}. The previous transformation implies
\begin{eqnarray}
&&\widetilde G^+=(C+D{\cal N})F^+=
(C + D{\cal N})(A+B{\cal N})^{-1} \widetilde F^+ \\
&&\rightarrow \mbox{
\parbox[t]{4.8cm}
{\fbox{$\widetilde{\cal N} = (C + D{\cal N})(A+B{\cal N})^{-1}$}}}
\label{tilNN}\end{eqnarray}
As remarked above, this tensor should be symmetric:
\begin{eqnarray}
 \rightarrow && (A+B{\cal N})^T  (C + D{\cal N})= (C + D{\cal N})^T
 (A+B{\cal N})
  \end{eqnarray}
which for a general ${\cal N}$ implies
\begin{equation}
A^T C-C^T A=0\ \ ,\ \ B^T D- D^T B=0\ \ , \ \  A^T D-C^T B=\unity\ .
\end{equation}
These equations express that ${\cal S}\in Sp(2m,\Rbar)$:
\begin{equation}
  {\cal S}^T  \Omega   {\cal S}   =  \Omega  \qquad\mbox{where}\qquad
\Omega=\pmatrix{0&\unity \cr -\unity &0\cr} \ .
\end{equation}
Some remarks are in order. First, these transformations act on
the field strengths. They generically rotate electric into magnetic fields
and vice versa. Such rotations, which are called duality
transformations, because in four space-time dimensions electric
and magnetic fields are dual to each other in the sense of
Poincar\'e duality, cannot be implemented on the vector potentials,
at least not in a local way. Therefore, the
use of these symplectic transformations is only legitimate for zero
gauge coupling constant. From now on, we deal
exclusively with Abelian gauge groups. Secondly, the Lagrangian is
not an invariant if $C$ and $B$ are not zero:
\begin{equation}
\Im \widetilde\cF^{+\Lambda}  \widetilde G_{+\Lambda}   =
  \Im \left(\cF^{+}  G_{+ }\right)
+\Im \left(2 \cF^{+} (C^T B) G_+
 + \cF^{+ }(C^T A) \cF^{+}
 +G_{+ } (D^T B) G_{+ } \right)\ .
\end{equation}
If $C\neq 0, B=0$ it is invariant up to a four--divergence,
as $\Im {\cal F}^+{\cal F}^+=-\ft i4 \epsilon^{\mu\nu\rho\sigma}
{\cal F}_{\mu\nu}{\cal F}_{\rho\sigma}$ and the matrices are real. Thirdly,
the transformations can also act on dyonic solutions of the field
equations and the vector $\pmatrix{q^\Lambda_m\cr q_{e\,\Lambda}\cr}$
of magnetic and electric charges transforms also as a symplectic
vector. The Schwinger-Zwanziger quantization condition restricts
these charges to a lattice (see also lectures of J. Harvey
\cite{lectHarvey}).
Invariance of this lattice restricts the symplectic
transformations to a discrete subgroup:
\begin{equation}
{\cal S}\in Sp(2m,\IZ) \ .
\end{equation}
Finally, the transformations with $B\neq 0$ will be
non--perturbative. This can be seen from the
fact that they do not leave the purely electric charges
invariant, or from the fact that \eqn{tilNN} shows that these
transformations invert ${\cal N}$, which plays the role of the gauge
coupling constant.
\subsection{Pseudo--symmetries and proper symmetries}
The transformations described above, change the matrix ${\cal N}$,
which are gauge coupling constants of the spin-1 fields. This
can be compared to diffeomorphisms of the scalar manifold
$z\to \hat z(z)$ which change the metric (which is the coupling
constant matrix for the kinetic energies of the scalars) and
${\cal N}$:
\[
\hat g_{\alpha\beta} (\hat z(z)) {\partial\hat z^\alpha
\over \partial  z ^\gamma}
{\partial\hat z ^\beta\over \partial  z ^\delta}= g_{\gamma\delta} (z)\ ;
\qquad \hat {\cal N}(\hat z(z))={\cal N}(z)\ .
\]
Diffeomorphisms and symplectic reparametrizations
are {\elevenit `Pseudo--symmetries':} \cite{christoi}
\begin{equation}
  D_{pseudo}=Diff({\cal M})\times Sp(2m,\Rbar)\ .
\end{equation}
They leave the action form invariant, but change the coupling
constants and thus are not invariances of the action.

If $\hat g_{\alpha\beta}(z)=g_{\alpha\beta}(z)$ then the
diffeomorphisms become
isometries of the manifold, and proper symmetries of the scalar
action. If these isometries are combined
with symplectic transformations such that
\begin{equation}
\hat {\widetilde{\cal N}}(z)={\cal N}(z)\ ,
\end{equation}
then this is a {\elevenit proper symmetry}. These are
invariances of the equations of motion (but not
necessarily of the action as not all transformations can be
implemented locally  on the gauge fields).
To extend the full group of isometries of the
scalar manifold to proper symmetries, one thus has to embed this
isometry group in $Sp(2m;\Rbar)$. In general, but not always
\cite{brokensi}, this seems to be realized in supersymmetric theories.

The simplest case is with one abelian vector. Then ${\cal N}=S$ is a
complex field, and the action is
\[
{\cal L}=\ft14(\Im S){\cal F}_{\mu\nu}{\cal F}^{\mu\nu}-\ft i8 (\Re S)
\epsilon^{\mu\nu\rho\sigma}{\cal F}_{\mu\nu} {\cal F}_{\rho\sigma}\ .
\]
The set of
Bianchi identities and field equations is invariant under symplectic
transformations, transforming the field $S$ as
\[ \widetilde S=\frac{C+DS}{A+BS}\qquad \mbox{where} \qquad AD-BC=1\ .
\]
If the rest of the action, in particular the kinetic term for $S$,
is also invariant under this transformation, then this
is a symmetry.  These transformations form an $Sp(2;\Rbar)=SL(2,\Rbar)$
symmetry. The $SL(2,\Zbar)$ subgroup is generated by
\begin{eqnarray}
\pmatrix {1&0\cr 1&1\cr}&&\pmatrix{0&1\cr -1&0\cr}\nonumber\\
\widetilde S=S+1 && \widetilde S=-\frac{1}{S}\ .
\end{eqnarray}
Note that $\Im S$ is invariant in the first transformation, while
the second one replaces $\Im S$ by its inverse. $\Im S$ is the coupling
constant. Therefore, the second transformation, can not be a
perturbative symmetry. It relates the strong and weak coupling
description of the theory.

For another example, namely $S$ and $T$ dualities in this
framework, see \cite{trspring}.

\subsection{Symplectic transformations in $N=2$} \label{ss:symplN2}
I now come back to rigid $N=2$ supersymmetry. There are $n$ vectors,
and the indices $\Lambda$ of the general theory are now the
$A$-indices. In the formulas of section~\ref{ss:rigidN2} the scalars,
$z^\alpha$ in the general theory, are the $X^A$.
The matrix $\bar {\cal N}_{AB}$ is now a well defined function. It
was given in \eqn{KFX} as
\begin{equation}
\bar {\cal N}_{AB} =F_{AB}=\frac{\partial F_A}{\partial X^B}\ .
\label{cNrigidF}\end{equation}
The last expression shows how we can obtain the transformation
\eqn{tilNN} for ${\cal N}$. Indeed, identify \cite{DWVP}
\begin{equation}
V\equiv \pmatrix{X^A\cr F_A\cr }\ , \label{Vrigid}
\end{equation}
as a symplectic vector, i.e. transforming under $Sp(2n,\Rbar)$ as
 $({\cal F}_{\mu\nu}^A,G_{A\mu\nu})$ in \eqn{FGsympl}:
\begin{eqnarray}
\tilde X^A &=& A^A_{\ B}X^B+B^{AB}F_B\ ,\nonumber\\
\tilde F_A &=& C_{AB} X^B +D_A^{\ B} F_B \ .
\label{symplXFX}\end{eqnarray}
This leads to
\begin{equation}
\widetilde{\bar {\cal N}}_{AB} =
\frac{\partial\widetilde F_B}{\partial \widetilde X^A}=
\left(C+D \bar {\cal N}\right) _{BC}
\frac{\partial X^C}{\partial \widetilde X^A}=
\left[\left(C+D \bar {\cal N}\right) \left(A+B \bar {\cal
N}\right)^{-1}\right]_{BA}\ ,  \label{tilcNrigid}
\end{equation}
which is the transformation we want for ${\cal N}$. So by identifying
$V$ as a symplectic vector, the structural relation \eqn{cNrigidF} is
preserved by \eqn{tilNN}. One may wonder whether \eqn{symplXFX} is
consistent with the definition that $F_A$ is the derivative of a
scalar function $F$. This requires that $\tilde F_A$ is the derivative
of a new function $\tilde F(\tilde X)$ w.r.t. the $\tilde X^A$.
The integrability condition
for the existence of $\tilde F$ can be seen from \eqn{tilcNrigid} to
be the condition that $\tilde{\cal N}$ is symmetric. We saw already
in section~\ref{ss:dualvector} that this is just the condition that
the transformation is symplectic.

Note that the argument for the existence of $\tilde F$ only applies
if the mapping $X^A\to \tilde X^A$ is invertible, such that
the $\tilde X^A$ are again independent coordinates. Hence, we need
that
\begin{equation}
\frac{\partial\tilde X^A}{\partial X^B}=  A^A_{\ B}+B^{AC}F_{CB}(X)
\label{invertApBF}
\end{equation}
is invertible (the full symplectic matrix is always
invertible). This we should anyway demand as $F_{CB}=\bar {\cal
N}_{CB}$, and the inverse of this matrix appears thus already from
the very beginning in \eqn{tilNN}. I put some emphasis on this point,
because this can be violated in supergravity.

As argued in the general theory, the transformations induced by
\eqn{tilNN} should extend to the other parts of the action. From
\eqn{KFX} it is clear that the \Ka\ potential is a symplectic invariant.
The fermionic sectors were checked in \cite{DWVP,ssss,f0art}.

Hence we obtain a new formulation of the theory, and thus of the
target-space manifold, in terms of the function $\tilde F(\tilde X)$.

We have to distinguish two  situations:\\
1. The function $\tilde F(\tilde X)$ is different from $F(\tilde X)$,
even taking into account \eqn{FapprF}.
In that case the two functions describe
equivalent classical field theories. We have a {\elevenit pseudo symmetry}.
These transformations are
called symplectic reparametrizations \cite{CecFerGir}.
Hence we may find a variety of descriptions of the same theory
in terms of different functions  $F$. \\
2. If a symplectic transformation leads to the same function $F$
(again up to \eqn{FapprF}), then we are dealing with a {\elevenit proper
symmetry}. As explained above, this invariance reflects itself in
an isometry of the target-space manifold.
Henceforth these symmetries are called
`duality symmetries', as they are generically accompanied by
duality transformations on the field equations and the Bianchi
identities.

E.g. the symplectic transformations with
\begin{eqnarray}
{\cal S}=\pmatrix {\unity & 0 \cr C &\unity }     \label{10C0}
\end{eqnarray}
do not change the $X^A$ and give $\tilde F=F+\ft12 C_{AB}X^AX^B$.
So these give proper symmetries for any symmetric matrix $C_{AB}$.
The symmetry of $C$ is required for ${\cal S}$ to be symplectic. In
the quantum theory $C$ will be restricted.
\pagebreak[2]
\subsection{Example: Perturbative duality symmetries
of the Seiberg-Witten model}
\label{ss:ex1loopSW}  \nopagebreak[4]
The  1-loop theory of the Seiberg-Witten model \cite{SeiWit}
gives a non-trivial example.
As classical theory Seiberg and Witten
took the $SU(2)$ theory \eqn{FSU2}. I mentioned already that the
potential is flat in one direction. The one-loop contributions give a
'quantum theory' for the massless field $X\equiv X^3$.
This has been calculated in
\cite{DVMNP}. It leads to an effective theory with
\begin{equation}
F(X)= \frac{i}{2\pi}X^2 \log \frac{X^2}{\Lambda^2}\ ,\label{asintpert}
\end{equation}
where $\Lambda$ is the renormalization mass scale.
Consider now the transformation defined by
\begin{equation}
{\cal S}=\pmatrix{-1 & 0\cr -2&-1\cr} \ .   \label{cSmonoinf}
\end{equation}
This leads to  (I write $F_A$ for the derivative w.r.t. the one
variable $X$)
\begin{eqnarray}
\widetilde X&=& -X= Xe^{i\pi}\nonumber\\
\widetilde F_A &=& -2 X-F_A(X)\nonumber\\
&=& -2X - \frac i\pi X \left(  \log \frac{X^2}{\Lambda^2}+1\right)
=F_A (\widetilde X(X))\ . \label{calcmonoinf}
\end{eqnarray}
Therefore this transformation leaves the function $F$ invariant and
is thus another (apart from \eqn{10C0}) duality symmetry.
This transformation corresponds to going around the singular point
$X=0$ for the square of $X$, which corresponds to the Casimir of the
original theory. In this way we cross the branch cut of the function
$F$, and this transformation is thus a `monodromy', as will be
explained in section~\ref{ss:dualSW}. We thus
find that the perturbative duality group is generated by
\begin{equation}
 \pmatrix{-1 & 0\cr -2&-1\cr} \qquad\mbox{and}\qquad
\pmatrix{1 & 0\cr 1&1\cr}\ . \label{pertmonoSW}
\end{equation}
A non-renormalization theorem says that in
$N=2$ there are no perturbative corrections beyond 1 loop.
\sectionsub{Symplectic formulation of rigid special geometry}
\label{ss:symplrigidsg}
I mentioned already at the end of section~\ref{ss:rigidN2} that one
can formulate the rigid special geometry in a reparametrization
invariant way \cite{FerStroCand,CdAF,special,CDFLL,modssym,prtrquat}.
In the previous section we saw that the symplectic group plays an
important role. The formulations which I will present below are
reparametrization invariant and
manifestly symplectic covariant. I give several equivalent
formulations, which are appropriate for making the connection to the
moduli space of Riemann surfaces in section~\ref{ss:RS}
\pagebreak[2]
\subsection{Coordinate independent description of rigid special
geometry} \label{ss:covrigidsg}
\nopagebreak[4]
The \Ka\ space is now parametrized by some holomorphic
coordinates $z^\alpha$ with $\alpha=1,...,n$.
The variables $X^A$ from above are holomorphic functions of the $z$.
But, as announced, I want to formulate it in a symplectic invariant
way, so I just take the symplectic vector \eqn{Vrigid} to be a
function of the $z^\alpha$, or, in other words, I define $n$
symplectic sections $V(z)$. I mentioned already that the \Ka\
potential is symplectic invariant. I rewrite now \eqn{KFX} as
\begin{equation}
K(z,\bar z)=i\langle V(z),\bar V(\bar z)\rangle \equiv iV^T(z)
\Omega\bar V(\bar z)\ ,
\end{equation}
where I defined a symplectic inner product. The metric is then
\begin{equation}
g_{\alpha\bar \beta}(z,\bar z)=
\partial_\alpha\partial_{\bar\beta }K(z,\bar z)=
i\langle U_\alpha(z),\bar U_{\bar \beta}(\bar z)\rangle \ ,
\label{gsympl}
\end{equation}
where I defined
\begin{equation}
U_\alpha \equiv \pmatrix{\del_\alpha  X^A \cr\del_\alpha  F_A\cr}
\equiv \pmatrix{ e_\alpha ^A \cr h_{A\alpha} \cr}\ . \label{defUalpha}
\end{equation}
One restriction which I mentioned already is the holomorphicity,
which can be expressed as
\begin{equation}
\partial_{\bar \beta}U_\alpha=0\ ;\qquad\partial_\beta\bar
U_{\bar \alpha}=0\ .\label{holU}
\end{equation}
In section~\ref{ss:rigidN2} the coordinates were the $X^A$. These are
now called 'special coordinates'.
So special coordinates are the ones where $z^\alpha=X^A$, or
$e^\alpha_A=\delta^\alpha_A$.

This does not yet contain the full definition of rigid special
geometry. Indeed, nothing of the above corresponds to the fact that
the $F_A$ are in special coordinates the derivatives of a function
$F$. I also did not yet give the matrix ${\cal N}_{AB}$ in general
coordinates.

I have $2n$ coordinates in my symplectic vector, which depend on $n$
coordinates $z^\alpha$. In the special coordinates the $F_A$ are
functions of the $X^A$, and I argued in section~\ref{ss:symplN2} that
symplectic transformations leave these as independent coordinates.
Therefore we can write
\begin{equation}
\partial_\alpha F_A = \frac{\partial F_A}{\partial
X^B}\partial_\alpha X^B\ .
\end{equation}
The matrix relating the lower to the upper components in \eqn{defUalpha}
is thus the one which is $\bar {\cal N}_{AB}$ in special coordinates.
I take this as definition of ${\cal N}$ in general coordinates:
\begin{equation}
h_{A\alpha} \equiv \bar {\cal N}_{AB} e_\alpha ^B\ . \label{defNgenrig}
\end{equation}
I stressed already in the general theory that this matrix ${\cal N}$
should be symmetric. It is clear here that this is equivalent to the
integrability condition for the existence of the scalar function $F$.
So therefore, the main equation defining special geometry is the
symmetry requirement of ${\cal N}$. Multiplying the above equation by
$e_\beta^A$, this condition is equivalent (as in these
arguments we assumed $e_\beta^B$ to be invertible) to
\begin{equation}
e^A_\beta h_{A\alpha}- h_{A\beta} e^A_\alpha=\langle
U_\alpha,U_\beta\rangle = 0\ .\label{symcNsym}
\end{equation}
This is a symplectic invariant condition. This condition together
with \eqn{gsympl} and \eqn{holU} {\elevenit define the rigid special
geometry.}
\subsection{Alternative definition in matrix form and curvature}
I still want to rewrite this definition in a matrix form which will
be useful for comparing with the moduli of Riemann surfaces.
First I rewrite \eqn{gsympl} and \eqn{symcNsym} as the matrix
equation
\begin{equation}
{\cal V}\Omega{\cal V}^T =-i\Omega\ ,\label{cVsympl}
\end{equation}
where ${\cal V}$ is the $2n\times 2n$ matrix
\begin{equation}
 {\cal V}\equiv
\left( \begin{array}{c}
 U_\alpha ^T \\[1mm] \bar U^{\alpha\,T}
  \end{array}\right)
 \equiv
\left( \begin{array}{cc}
e_\alpha^A &h_{A\alpha}\\[1mm]
 g^{\alpha \bar\beta}\bar e^A_{\bar \beta} &g^{\alpha \bar\beta}
\bar h_{A\bar \beta}\end{array}\right) \ ,   \label{defcVrigid1}
\end{equation}
and $\bar U^{\alpha} \equiv  g^{\alpha \bar\beta}
 \bar U_{\bar \beta}$. The result
\eqn{cVsympl} states that $\cal V$ is isomorphic to a symplectic
matrix. Therefore
${\cal V}$ is invertible and we can define $Sp(2n)$ connections
${\cal A}_\alpha$ and ${\cal A}_{\bar \alpha}$ such that
\begin{equation}
{\cal D}_\alpha{\cal V}= {\cal A}_\alpha{\cal V}\,, \qquad
{\cal D}_{\bar\alpha} {\cal V}={\cal A}_{\bar \alpha}{\cal V}\,.
\label{flatness}
\end{equation}
Here we introduced covariant derivatives where the Levi-Civita
connections\footnote{In general for \Ka\ manifolds, the
connection is only non--zero when all indices are holomorphic or
antiholomorphic (${\cal D}_\alpha U_\beta=\partial_\alpha U_\beta
-\Gamma_{\alpha\beta}^\gamma U_\gamma$). In the curvature tensor
with all indices
down, 2 should be holomorphic and 2 antiholomorphic.
The nonzero components are $\Gamma_{\alpha \beta }^\gamma  =
g^{\gamma \bar \delta }
\partial_\beta  g_{\alpha \bar \delta }$, and
$R^\gamma{}_{\alpha \beta\bar \delta} =\partial_{\bar \delta}
\Gamma_{\alpha \beta }^\gamma  $ and all those
related by the symmetries of the curvature tensor.}
appear because there are $\alpha$ indices hidden in ${\cal V}$:

The values of these two connections can be computed from multiplying
the above equation with $\Omega {\cal V}^T$:
\begin{equation}
-{\cal A}_\alpha i\Omega={\cal D}_\alpha{\cal V}\Omega{\cal V}^T=
\pmatrix{\langle{\cal D}_\alpha U_\beta,U_\gamma\rangle &
\langle{\cal D}_\alpha U_\beta, \bar U^\gamma\rangle \cr
\langle{\cal D}_\alpha \bar U^\beta,U_\gamma\rangle &
\langle{\cal D}_\alpha \bar U^\beta, \bar U^\gamma\rangle \cr}
\end{equation}
By \eqn{holU} and the covariant derivative of \eqn{gsympl} we know
already that only the upper left component of the last matrix can be
non-zero:
\begin{equation}
C_{\alpha\beta\gamma}\equiv
-i\langle{\cal D}_\alpha U_\beta,U_\gamma\rangle \ .  \label{defC}
\end{equation}
The covariant derivative on \eqn{symcNsym} implies that
$C_{\alpha\beta\gamma}$ is symmetric in $(\beta\gamma)$. Remembering the
definition \eqn{defUalpha} we know that it is symmetric in all
indices. We find
\begin{equation}
 {\cal A}_\alpha =\pmatrix{0& C_{\alpha \beta \gamma } \cr 0&0\cr}\
 ;\qquad
 {\cal A}_{\bar \alpha} =\pmatrix{0&0 \cr \bar C_{\bar\alpha}{}^{\beta
 \gamma}&0\cr}\ .   \label{defcArigid}
\end{equation}

Later we want to use the matrix ${\cal V}$ as a starting point. So I
give now the definition of special geometry from this point of view.
We then start with a square matrix ${\cal V}$ whose components are
denoted as
\begin{equation}
{\cal V}\equiv\pmatrix{e_\alpha^A &h_{A\alpha}\cr
\bar e^{\alpha A} & \bar h^\alpha_A\cr}\ .
\end{equation}
Here, the
requirements for special geometry are \eqn{cVsympl} and \eqn{flatness}
with  \eqn{defcArigid} in which $C$ is a symmetric tensor. Although
the metric is not yet given, the connection is  determined by the
equations themselves. Indeed, the upper left component of \eqn{flatness}
multiplied by $e_A^\alpha$, the inverse of $e^A_\alpha$, gives
\begin{equation}
 \Gamma_{\alpha\beta}^\gamma= \hat \Gamma_{\alpha \beta}^\gamma -
\bar e^{\gamma A} C_{\alpha\beta\delta}e^\delta_A \qquad\mbox{with}
\qquad   \hat \Gamma_{\alpha \beta }^\gamma =e^\gamma _A
\partial_\beta e_\alpha ^A \ .\label{Gammahat}
\end{equation}

The differential equations give a lot of information.
The conditions (\ref{flatness}) imply that the combined
connection consisting of $\cal A$ and the Levi-Civita
connections must be flat. The integrability conditions are
\begin{equation}
\left[{\cal D}_\gamma -{\cal A}_\gamma ,{\cal D}_{\bar \delta }
-{\cal A}_{\bar \delta } \right]{\cal V}=0\ ;\qquad
\left[{\cal D}_\gamma -{\cal A}_\gamma ,{\cal D}_\delta -
{\cal A}_\delta \right]{\cal V} =0\ , \label{integrcA}
\end{equation}
The upper component of the first one is
\begin{equation}
U_\beta R^\beta{}_{\alpha\gamma\bar \delta}+\bar U^\beta{\cal
D}_{\bar \delta}C_{\gamma\alpha\beta}+C_{\gamma\alpha\epsilon}
\bar C_{\bar \delta}{}^{\epsilon\beta}U_\beta=0 \ ,
\end{equation}
where I used the definition of the curvature tensor
\begin{equation}
\left[{\cal D}_\gamma ,{\cal D}_{\bar \delta }\right] U_\beta =
U_\alpha  R^\alpha {}_{\beta \gamma \bar \delta } \ .
\end{equation}
Using the symplectic orthogonalities \eqn{gsympl} and \eqn{symcNsym}
this equation can be split. The first relation we obtain is that the
Riemann curvature is given by
\begin{equation}   R^\alpha {}_{\beta \gamma}{}^\delta
=- C_{\beta \gamma \epsilon}
\bar C^{\alpha\delta\epsilon} \ . \label{Rrigid}
\end{equation}
Secondly, the tensor $C_{\alpha \beta \gamma }$ satisfies the following
two conditions (the second one follows from the second condition
in \eqn{integrcA})
\begin{equation}
{\cal D}_{\bar \alpha }C_{\beta \gamma \delta } ={\cal D}_{[\alpha }
C_{\beta ]\gamma \delta } = 0 \ . \label{diffeqnC}
\end{equation}
The last one implies that $C_{\alpha \beta \gamma }$ can be written
as the third covariant derivative of some scalar function. To make
the connection with section~\ref{ss:rigidN2},  \eqn{defC} can
be worked out using \eqn{defNgenrig} and \eqn{KFX} to obtain
\begin{equation}
C_{\alpha \beta \gamma }=  i e_\alpha ^A e_\beta ^B e_\gamma ^C
F_{ABC} \ .
\end{equation}
\subsection{Holomorphic equations}
Apart from the coordinate invariance, the above formulation is also
invariant under gauge transformations \cite{CDFLL}
\begin{equation}
{\cal V}'= S^{-1}\, {\cal V}  \ ;\qquad {\cal A}'_\alpha=S^{-1}
\left({\cal A}_\alpha-{\cal D}_\alpha \right) S \ ;\qquad {\cal A}'
_{\bar \alpha}=S^{-1}
\left({\cal A}_{\bar \alpha}-{\cal D}_{\bar \alpha}\right) S
\label{transfS}\end{equation}
for $S$ a symplectic element: $S\Omega S^T=\Omega$. Such a
transformation can be used to reduce the equations to holomorphic
ones.

Indeed, perform the symplectic transformation
\begin{equation}
S= \pmatrix{\delta_\alpha^\beta & 0\cr
\bar e^{A(\alpha}e^{\beta)}_A &\delta_\beta^\alpha\cr }\ ,
\end{equation}
which gives
\begin{equation}
{\cal V}'=\pmatrix{e_\alpha^A&h_{\alpha A}\cr f^{\alpha\beta}e_\beta^A
& \bar h'^\alpha_A\cr}\ .
\end{equation}
The value of $\bar h'$ does not matter, but it is important to note
that $f^{\alpha\beta} $ is antisymmetric  (its value is
$f^{\alpha\beta}\equiv\bar e^{A[\alpha}\,e^{\beta]}_A $). This is
sufficient to obtain from the constraints \eqn{cVsympl} that
\begin{equation}
 f^{\alpha\beta} =0\ ;\qquad \bar h'^\alpha_A =-i\,e^\alpha_A
 \qquad\Rightarrow\ {\cal V}'=\pmatrix{e_\alpha^A&h_{\alpha A}\cr
0&-i\,e^\alpha_A \cr} \ .     \label{holcVrigid}
\end{equation}
The remaining part of \eqn{cVsympl} is \eqn{symcNsym}.

For the differential equations, consider first ${\cal A}'_{\bar \alpha}$.
{}From \eqn{transfS} it follows that only the lower left component is
non--zero. However, then \eqn{flatness} with \eqn{holcVrigid} imply
that this component should also be zero, leaving us with
\begin{equation}
{\cal D}_{\bar \alpha}{\cal V}'=\partial_{\bar \alpha}{\cal V}'= 0\ .
\end{equation}
So here the holomorphicity becomes apparant. Remains the calculation of
${\cal A}' _\alpha$. Again  \eqn{flatness} and \eqn{holcVrigid} imply
that the lower left component is zero. Further it is easy to check
that the upper right component does not change. It is then sufficient
to calculate the upper left component, and the lower right follows
from consistency:
\begin{equation}
{\cal A}'_\alpha=\pmatrix{X_{\alpha\beta}^\gamma & C_{\alpha\beta\gamma}
\cr 0& -X_{\alpha\gamma}^\beta \cr} \qquad\mbox{with}\qquad
X_{\alpha\beta}^\gamma  =
C_{\alpha\beta\delta}\,\bar e^{A(\delta}\,e^{\gamma)}_A
\end{equation}
(The symmetrization symbol in the expression of $X$ is superfluous in
view of $f^{\alpha\beta}=0$).
If one writes the
differential equation with an ordinary rather than a covariant
derivative, then the diagonal elements simplify:
\begin{equation}
\left(\partial_\alpha  -\widehat {\cal A}_\alpha \right){\cal V}'
=0 \qquad\mbox{with} \qquad\widehat  {\cal A}_\alpha  =
\pmatrix {\hat \Gamma_{\alpha \beta }^\gamma &
C_{\alpha \beta \gamma }  \cr 0 & - \hat \Gamma_{\alpha \gamma }^\beta }
\ ,   \label{flatnesshat}
\end{equation}
and $\hat \Gamma $, which in facts follows from this equation, given in
\eqn{Gammahat}. So in this formulation special geometry is determined by
\eqn{flatnesshat} on the holomorphic matrix \eqn{holcVrigid}, which
moreover should satisfy \eqn{symcNsym}.

The differential equations can be combined to a second order
equation.
E.g. for $n=1$, if we write the upper components of ${\cal V}'$
generically as $f(z)$ and the lower component as $g(z)$, the
differential equations are
\begin{equation}
(\partial -\hat \Gamma)f  + C g =0 \ ;\qquad
(\partial +\hat \Gamma)g =0 \ .
\end{equation}
Combining these we have  for $f$:
\cite{modssym}
\begin{equation}
 (\partial +\hat \Gamma) C^{-1}(\partial -\hat \Gamma)f =0\ .
\label{2ndoden1}\end{equation}

\sectionsub{Special geometry in the moduli space
of Riemann surfaces}\label{ss:RS}
After identifying objects in $N=2$ supersymmetry with objects in the
moduli space of Riemann surfaces, I will show that there are
differential equations (Picard-Fuchs equations) which are related to
the equations defining the special geometry. These will be solved to
lead to a symplectic section compatible with the constraints of
special geometry. As explained in the introduction, this symplectic
section is conjectured to yield the full perturbative and
non-perturbative quantum theory of some models. In particular, I
will explain the Seiberg-Witten model from this point of view and
discuss its duality symmetries.
\subsection{The period matrix}
\label{ss:moduliRS}
The matrix ${\cal V}$ of special geometry will be identified with the
period matrix of Riemann surfaces
(up to a gauge transformation \eqn{transfS} on the
left and a symplectic constant transformation on the right).
This will depend on the moduli
which define deformations of the surface. These moduli will be
identified with the scalars of special geometry. In this first
subsection, the dependence on the moduli will not yet be essential.
That will be the subject of section~\ref{ss:PicardFuchs}.

The period matrix is
\begin{equation}
\tilde{\cal V}= \pmatrix{\tilde U_\alpha ^T \cr \tilde {
\bar U}{}^{\alpha \, T}\cr}=\pmatrix {\int_{A^A} \,
\gamma_{\alpha }& \int_{B_B}\gamma_{\alpha }\cr \int_{A^A} \,
\bar \gamma^{\alpha }& \int_{B_B}\bar \gamma^{\alpha }\cr}\ ,
\label{defcalVRiemann}\end{equation}
where $\gamma_\alpha$ is a basis of $H^{1,0}$, i.e. the
closed and non--exact one-forms, while $\bar \gamma^\alpha$ are a
basis of $(0,1)$ forms. On a genus $n$ surface
there are $n+n$ such forms. These are integrated  in \eqn{defcalVRiemann}
along a canonical homology basis of 1-cycles with intersection numbers:
\begin{equation}
A_A \cap A_B = 0\ ; \qquad \quad B^A \cap B^B = 0 \ ; \qquad
A_A \cap B^B= -B^B \cap A_A = \delta_A^B\ . \label{rie_4}
\end{equation}
The relation \eqn{cVsympl} follows now from the identity for two
1-forms $\lambda$ and $\chi$
\begin{equation}
\sum_{A}\left[\int_{A^A} \lambda \cdot \int_{B_A} \chi -
\int_{B_A} \lambda \cdot \int_{A^A} \chi  \right]=\int\int  \lambda
\wedge \chi  \ ,
\end{equation}
where the integral at the right hand side is over the full Riemann
surface. The product of two (1,0) forms is then obviously zero, and
the basis of (0,1)-forms can be chosen then such that \eqn{cVsympl}
is satisfied.

The second equation, \eqn{flatness}, will be the subject of the next
subsection. First, I introduce a description of the Riemann surface
(with its moduli) as an elliptic curve, and give an expression for the
elements in $\tilde{\cal V}$. The formulations given here
\cite{monodrcy} are chosen
such that they can easily be generalized to the Calabi-Yau case.
I use a
3-dimensional weighted projective complex space $(Z, X, Y)$. The
surface is defined as the points of this space where a
pseudo-homogenous holomorphic polynomial vanishes. Due to the
homogeneity we can view this in the weighted projective space, as
a~1 complex dimensional surface, which is the Riemann surface. For
genus~1, which will become the main example, I use the polynomial
\begin{equation}
0~=~W(X, Y, Z; u)~=~ -Z^2\, + \, \frac {1}{4} \, \left ( X^4 + Y^4
\right ) \, + \, \frac{u}{2} X^2 Y^2 \ ,  \label{cWSW}
\end{equation}
in the projective space where $Z$ has weight~2, and $X$ and $Y$ have
weight~1. There is one complex parameter (modulus) $u$. The holomorphic
function $W$ can also be seen as a Landau-Ginzburg superpotential,
but I will not use that connection here.

To obtain the forms, we make use of the Griffiths mapping
\cite{griffiths}, which relates the holomorphic forms to elements of
the chiral ring
\begin{equation}
{\cal R} ({\cal W} )~{\stackrel{\rm def}{=}}~
 \Cbar [X,Y,Z]/\partial {\cal W}\ ,  \label{chiralring}
\end{equation}
i.e. polynomials of a certain degree in the variables $X,Y,Z$, where
the derivatives of ${\cal W}$ are divided out. The 1-forms on the
Riemann surface ${\cal M}$ are represented as
\begin{equation}
\left. \gamma\right|_{\cal M}=\int_\Gamma \frac{\gamma\wedge
d{\cal W}}{{\cal W}}=\int_\Gamma\frac{P^\gamma_{k|\nu}(X,Y,Z)}
{{\cal W}^{k+1}} \omega\ ,      \label{Griffithgamma}
\end{equation}
(everything can depend on the moduli $u$)
where $\Gamma$ is a 1-cycle around the surface ${\cal M}$
in the 4-dimensional space (6-dimensions from the complex $X,Y,Z$ $-$
2 from homogeneity). The first equality is based on a generalisation of
the residue theorem. In the second equality exact forms were
discarded because of the integration over a cycle
(see e.g. \cite{fresoriabook} sect. 5.9 for more details).
$P^\gamma_{k|\nu}(X,Y,Z)$ is a pseudo homogeneous polynomial of a
degree such that \eqn{Griffithgamma} has weight
zero. That degree is denoted as $k|\nu$, where $\nu$ is the degree
of ${\cal W}$). In the example
\eqn{cWSW} we have $\nu=4$, and we will see that $\omega$ has degree~4,
so that
\begin{equation}
k|\nu =k|4= 4k\ . \label{kbarnu}
\end{equation}
  The 'volume form' $\omega$ is \cite{Morrison}
\begin{equation}
\omega=2(X\,dY \wedge dZ +Y\, dZ\wedge dX +2\,Z\, dX\wedge dY) \ .
\end{equation}
The important property of $\omega$ is that for any meromorphic
function ${\cal P}^\Lambda(X,Y,Z)$
\begin{equation}
\omega\,\partial_\Lambda {\cal P}^\Lambda(X,Y,Z) = \dop\Theta
\label{propomega}
\end{equation}
for some form $\Theta$ if the degrees of $\partial_\Lambda
{\cal P}^\Lambda$ and $\omega$ add up to zero
(I introduced the notation $X^\Lambda$ for $X,Y,Z$). The proof of
this statement is given in appendix~\ref{app:volform} for any
pseudo-homogeneous space. In fact,
all the above can be generalised for more variables (see below for
Calabi-Yau) and for manifolds defined by the intersection of the
vanishing locus of  several polynomia.

If $P^\gamma_{k|\nu}(X,Y,Z)$ has terms
proportional to $\partial_\Lambda {\cal W}$, then in \eqn{Griffithgamma}
they occur as proportional to $\partial_\Lambda {\cal W}^{-k}$.
So, using \eqn{propomega}, they can be removed by adding
a total differential to the integrand (which will decrease $k$
and the degree of the remaining $P$). The total differential is
integrated over a cycle and thus does not contribute. This shows that
for the polynomials $P^\gamma$, we suffice by a basis of the elements
of the ring \eqn{chiralring} with the correct weights. In the chiral
ring there is a maximal weight, which therefore restricts also $k$.
In general the upper limit of $k$ is the dimension (over $\Cbar$) of
the manifold. In our example in the chiral ring $X^3$, $Y^3$ and any
$Z$ terms can be removed in this way, and the maximal element is
thus $X^2Y^2$, i.e. of weight~4. Comparing with
\eqn{kbarnu}, the maximal $k$ is~1, as it should be on a Riemann
surface. In fact, in
general for manifolds of complex dimension $N$, the
polynomials $P^\gamma_{k|\nu}$ are associated to the elements in
\begin{equation}
H^{N,0}\oplus H^{N-1,1} \oplus  \cdots \oplus   H^{N-k,k}\ .
\end{equation}
In our case, we thus have that for $k=0$ the polynomial has degree~0,
and can thus just be a constant. It is associated with the unique
$(1,0)$ form on the genus~1 surface. For $k=1$ there is only
$X^2Y^2$, and together they are associated to all the 1-forms.
{\elevenit Exercise}: check that for genus $n$ surface, defined by ${\cal
W}=-Z^2+X^{2n+2}+Y^{2n+2}$ one obtains indeed $n$ (1,0) and $n$ (0,1)
forms.

We now integrate these forms over the $A$ or $B$ cycles to obtain
the elements of the period matrix. The product of that cycle with
$\Gamma$ in \eqn{Griffithgamma} is a 2-cycle which will be
generically denoted by $C$. So we obtain (I have not imposed here a
normalization condition as in \eqn{cVsympl})
\begin{equation}
\tilde{\cal V} = \left ( \matrix { \int_C \,  \frac{1}{{\cal W}(X,Y,Z;u)}
 \, \omega \cr
\int_C   \frac{X^2 \, Y^2 }{{\cal W}^2(X,Y,Z;u)}  \,\omega
 \cr } \right )\ .  \label{PeriodsSW}
\end{equation}
\pagebreak[2]
\subsection{Picard--Fuchs equations}    \label{ss:PicardFuchs}
The forms and periods depend on the moduli $u$. As we will see in
the example, the derivatives of this period matrix can be
re-expressed linearly in terms of the periods such that we get the
differential equations
\begin{equation}
\partial_u \tilde{\cal V}= \tilde A_u\tilde{\cal V}\ .\label{diedro_10}
\end{equation}
These differential equations are called
Picard-Fuchs equations. They are
similar to the equations \eqn{flatness} of rigid special \Ka\ manifolds, and
therefore complete the identification of special geometry with the
geometry of the moduli space of the Riemann surface. In
section~\ref{ss:solSW} we will see that these
differential equations allow us to calculate the period matrix.
The main tool to derive the Picard-Fuchs equations is \eqn{propomega}.

Higher genus surfaces have more than $n$ complex moduli. To make the
connection with special geometry, one then
has to define a surface depending on exactly $n$
moduli $u^\alpha$. The choice of these $n$ moduli is non-trivial
\cite{modulisun}.

I will concentrate on the case of a genus~1 surface which has one
complex modulus \cite{SeiWit}. If I take the derivative of
\eqn{PeriodsSW} with respect to $u$, then this derivative acts on
the denominator, and $\partial_u{\cal W}=\ft12 X^2Y^2$. For the
upper component, that is all: the derivative gives immediately
$(-\ft12)$ the lower component.
Taking the derivative of the lower component, one gets $X^4Y^4$
in the numerator, which is not in the chiral ring, and thus has to be
reduced, e.g.
\begin{equation}
(1-u^2)X^4Y^4 = X\,Y^4\partial_X{\cal W} -u\,X^2Y^3\partial_Y{\cal
W}\ .
\end{equation}
This gives
\begin{eqnarray}
(1-u^2)\partial_u
\int_C   \frac{X^2 \, Y^2 }{{\cal W}^2(X,Y,Z;u)}  \,\omega &=&
-(1-u^2)
\int_C   \frac{X^4 \, Y^4 }{{\cal W}^3(X,Y,Z;u)}  \,\omega \nonumber\\
&=& \ft12\int_C \omega\,X \, Y^4 \partial_X{\cal W}^{-2}
-\ft12 u \int_C \omega\, X^2Y^3\partial_Y {\cal W}^{-2} \nonumber\\
&=&- \ft12\int_C \omega \, Y^4 {\cal W}^{-2}
+\ft32u \int_C \omega\, X^2Y^2 {\cal W}^{-2}  \nonumber\\
&=& \ft12\int_C \omega \, Y \partial_Y {\cal W}^{-1}
+2\, u \int_C \omega\, X^2Y^2 {\cal W}^{-2}\ .
\end{eqnarray}
We thus obtain \eqn{diedro_10} where the $2\times 2$ matrix
connection $\tilde{\cal A}_u$ is given by:
\begin{equation}
\tilde{\cal A}_u = \left(\matrix{0 & -{1\over 2}
\cr {- 1/2\over 1 - u^2}  &
{2 u\over 1 - u^2} }\right)  \ .
\end{equation}
The differential equations corresponding to the 'A' and 'B'-cycles
are of course identical. For both I can combine the upper and lower
equations to a second order differential equation for the upper
component, which I denote generically by $f(u)$:
\begin{equation}
\label{diedro_11}
\biggl({\dop^2\over \dop u^2} - {2 u\over 1 - u^2} {\dop\over \dop u}
- {1\over 4} {1\over 1 - u^2} \biggr) f(u) = 0 \ .
\end{equation}
Comparing this equation with \eqn{2ndoden1}, we can already conclude
that $C\propto (1-u^2)^{-1}$, which determines the curvature
(up to a constant).
\pagebreak[2]
\subsection{Solutions} \label{ss:solSW}
Solving the differential equation \eqn{diedro_11} will lead us now to
the symplectic vector, determining again an $N=2$ rigid supersymmetric
theory. As mentioned in the introduction, that theory is conjectured
by Seiberg and Witten to be the full quantum theory for the modulus
$u$.

If we change variable, setting $w = (1 + u)/2$, \eqn{diedro_11}
becomes a hypergeometric equation
of parameters $a=1/2, b = 1/2, c=1$. The corresponding hypergeometric
functions are elliptic integrals. Some general facts about these are
given in appendix~\ref{app:ellint}. We choose  as two linear
independent solutions (writing, as in section~\ref{ss:ex1loopSW}, $F_A$
for the one component of that vector).
\begin{equation}
\label{periodando}  U_u=
\left( \begin{array}{l}
{\del_u X} \equiv f^{(1)}(u) =\frac{2}{\pi}\left[
K \left ({{1+u} \over 2} \right ) +
 {\rm i} \, K \left ({{1-u} \over 2} \right )\right]
= \frac{2}{\pi}
\sqrt{{2 \over {1+u}}}
\, K \left (
 {2 \over {1+u}} \right )  \\
\partial_u \,F_A \equiv
f^{(2)}(u) =  \frac{2}{\pi}
{\rm i} K \left ({{1-u} \over 2} \right )\end{array}\right)\ ,
\end{equation}
where we understand here and further $+i\epsilon$ for the argument
${2 \over {1+u}} $ on the positive real axis for the definition of $K$
as given in appendix~\ref{app:ellint}.
As it is obvious, $f^{(1)}(u)$ and $f^{(2)}(u)$ just provide a
basis of two independent solutions.
The reason why precisely $f^{(1)}(u)$ and $f^{(2)}(u)$ are
respectively
identified with $\partial_u X$ and
$\partial_u \,F_A$  is given
by the boundary conditions imposed at infinity. When
$ u \, \rightarrow \, \infty $, the
special coordinate $X(u)$ must approach the value it
has in the original microscopic $SU(2)$ gauge theory. The parameter
$u$ corresponds to a gauge--invariant quantity in the microscopic
theory, so it is associated to the gauge invariant polynomial
$X^1X^1+X^2X^2+X^3X^3$. This is now restricted to
the Cartan  subalgebra, i.e. $u$ is proportional to the square of
$X\equiv X^3$.
Correspondingly the boundary condition at infinity
for $X(u)$ is  (choosing a convenient normalisation)
\begin{equation}
X(u) \approx 2\, \sqrt{2u}+\ldots \qquad\mbox{for }u\to\infty\ .
\label{asintuno}
\end{equation}
At the same time when $ u \, \rightarrow \, \infty $ the
non perturbative rigid special geometry
must approach its perturbative limit  defined by the
prepotential \eqn{asintpert}.
Combining eq.(\ref{asintuno}) and (\ref{asintpert}) we obtain
\begin{equation}
F_A(u)\, \approx \, {{\rm i}\over {\pi}}
2\, \sqrt{2u} \, {\rm log } \, u +\ldots \qquad\mbox{for }
u\to\infty\ .
\label{asintdue}
\end{equation}
I now show that these
boundary conditions are realized by
the choice of  (\ref{periodando}). Integrating the latter,
using \eqn{intE1x} and \eqn{intK}, gives
\begin{equation}
\label{integrando} V=
\left( \begin{array}{l}
X(u) = \frac{2}{\pi}\int_{u_0}^u \,\sqrt{{2 \over {1+t}}} \, K \left (
 {2 \over {1+t}} \right ) \, dt \, = \,
 {8\over {\pi}} \, \sqrt{\frac{ 1+u}{2}} \,
 E \left ( {2 \over {1+u}} \right ) \, + \, {\rm const} \\
 F_A (u) =\frac{2}{\pi}
i \, \int_{u_0}^{u} \, K \left ({{1-t} \over 2} \right )
\, dt \, = -\frac{4i}{\pi} \,(1-u)\, B\left( \frac{1-u}{2}\right)
\, + \, {\rm const.}
\end{array}\right)
\end{equation}
Choosing zero for the integration constants, the result
(\ref{integrando}) coincides with the integral representations
originally given by Seiberg and Witten \cite{SeiWit}. Indeed, performing
the substitutions $y^2=(1-x)/2$ in the integral representation of the
first one, and $y^2(u-1)=x-1$ in the second, leads to
\begin{equation}
\label{paragonando} V=
\left(\begin{array}{l}
X(u) = 2a(u)=\frac{2\sqrt{2}}{\pi}
\int_{-1}^{1} \,\sqrt{{u-x} \over {1-x^2}} \, dx \\
F_A (u) =  2\,a_D(u)=2i\frac{\sqrt{2}}{\pi}
 \int_{1}^{u} \,
\sqrt{{u-x} \over {1-x^2}} \, dx
\end{array}\right)\ .
\end{equation}
It is clear that the first one leads to \eqn{asintuno}. For the
second one, a substitution $x=uz$ exhibits the asymptotic behaviour
as in \eqn{asintdue} \cite{SeiWit}.
\pagebreak[2]
\subsection{Duality symmetries}\label{ss:dualSW}
The duality symmetry group $\Gamma_D$ consists of the symmetries of
the potential ${\cal W}$ and the group generated by the monodromies
around singular points. All these are given by integer valued
symplectic matrices $\gamma \in \Gamma_D \subset Sp(2n, \ZZ)$ that
act on the symplectic vector $U_\alpha $. Given the geometrical
interpretation (\ref{defcalVRiemann}) of these vectors they
correspond to changes of the canonical homology basis respecting the
intersection matrix (\ref{rie_4}).

Let us start by the symmetries of the defining equation. The
symmetry group $\Gamma_{{\cal W}}$ can be defined by considering
linear transformations ${\elevenbf X} \, \to \, M_x(u){\elevenbf X}$ of the
quasi--homogeneous coordinate vector ${\elevenbf X}=(X, Y, Z)$.
For preservation of the weights $M_x(u)$ should be
block diagonal. $\Gamma_{{\cal W}}$ consists of
the transformations such that
\begin{equation}
{\cal W}(M_x(u){\elevenbf X};u)=f_x(u) {\cal W}({\elevenbf X} ;
u'_x(u))\qquad \mbox{and}\qquad
 \omega(M_x(u){\elevenbf X} ) =   g_x(u)\omega({\elevenbf X})\ ,
\label{diedro_2}
\end{equation}
where $u'_x(u)$ is a (generally non--linear) transformation of the
moduli and $f_x(u)$ and $g_x(u)$ are overall rescalings of the
superpotential and the volume form that depend both on the moduli
$u$ and on the chosen transformation $x$. In the supergravity case
all elements of $\Gamma_{{\cal W}}$ are duality symmetries. However,
here the symplectic vector $U_u$, the
upper component of \eqn{PeriodsSW}, satisfies
\begin{equation}
U_u(u'_x(u)) =\int_C \frac{f_x(u)}{{\cal W}(M_x(u){\elevenbf X};u)}
\frac{ \omega(M_x(u){\elevenbf X} )}{g_x(u)}=\frac{f_x(u)}{g_x(u)}
\, S_x \, U_u(u)\ .\label{rescfactorsU}
\end{equation}
In the last step, a change of integration variables was performed,
which transforms also the $A$ and $B$ cycles, leading to the
symplectic matrix $S_x$.
If $ \frac{f_x(u)}{g_x(u)} $ is a constant, then the right hand side
gives also a solution to the differential equations, and we have a
duality symmetry. Therefore, only
$\Gamma_{\cal W}^0 \subset \Gamma_{\cal W}$
given by the transformations that have a constant
$ \frac{f_x(u)}{g_x(u)} $ acts as an isometry group for the moduli
space \cite{monodrcy}.

For \eqn{cWSW}, the
symmetry group is $\Gamma_{\cal W}=D_3$ \cite{giveonlopez}
defined by the following generators and relations
\begin{equation}
{B}^2= \unity\quad , \quad C^3=\unity \quad, \quad
(CB)^2 =\unity
\end{equation}
with the following action on the homogeneous coordinates
and the modulus $u$ \footnote{The full group connects 6 points as in
\eqn{6ptsK}: $\pm u,\,\pm \frac{u-3}{u+1},\, \pm \frac{u+3}{u-1}$. }
\begin{equation}
\begin{array}{llll}
M_{B} =
\left (\matrix {i & 0&0 \cr 0 & 1&0 \cr 0&0&1\cr}
\right )\ ; & u'_{B} (u)\, = \, - u \ ; & f_{B}(u)\ ;\,
= \, 1& g_B=i\\
M_C = {\frac{1}{\sqrt{2}}} \,
\left (\matrix {i & 1&0  \cr -i & 1&0  \cr 0&0&\sqrt{1+u}\cr}
\right )\ ; & u'_C (u)\, = \,{\frac{u-3}{u+1}}\ ; & f_C(u)\, = \,
{\frac{1+u}{2}}\ ;&g_C=i\sqrt{\frac{1+u}{2}}\ .
\end{array}
\label{diedro_5}
\end{equation}
{}From the value of $f_x$ and $g_x$ we thus see that only the
$\ZZ_2$ cyclic group generated by $B$ is actually realized as an
isometry group of the rigid special K\"ahlerian metric.
The transformation $u\to -u$ acts on $U_u$ as
\begin{equation}
U_u(-u)=-i\pmatrix{-1&2\cr -1&1}U_u(u)\qquad
\Rightarrow \ B=-iS_B=-i \pmatrix{-1&2\cr -1&1}  \ . \label{defBR}
\end{equation}
This transformation has a physical interpretation
as R-symmetry. It is precisely the requested R-symmetry for the
topological twist \cite{rtwist}.

The monodromy group generators correspond to analytic continuation
of the solutions around the singular points and have a matrix
action on the solution two--vector $U_u \rightarrow T_{(Q)}U_u$,
where $Q$ denotes any of the singular values of $u$,
and $T_{(Q)}$ is a $2\times 2$ matrix. One can see that the
polynomial \eqn{cWSW} degenerates for $u=1,\, -1$ and $\infty$. Of
course, one can also find these singular points by considering the
solution vector \eqn{periodando}.
Under the action of the group $\Gamma_{\cal W}^0=\ZZ_2$
generated by $B :\,  u \to -u$, the singular
points fall in two orbits: $\{ 1,-1\}$ and $\{ \infty \}$.

Let us derive the structure of the monodromy group by
direct evaluation. The monodromies will be defined going around the
singularity in counter clockwise direction. The calculation of the
monodromy around $u=\infty$ is similar to \eqn{calcmonoinf}. The
matrix ${\cal S}$ \eqn{cSmonoinf} gives now
\begin{equation}
T_{(\infty)}= \pmatrix{-1&0\cr -2&-1} \ . \label{Tinfty}
\end{equation}
The monodromy around $u=1$ is found from \eqn{monoK} and
$K(1-y')=K(1-y)$:
\begin{equation}
U_u(1+r e^{2\pi i})=\pmatrix{1&-2\cr 0&1}
U_u(u)\qquad\mbox{for }0<r<2\qquad
\Rightarrow \ T_{(1)}= \pmatrix{1&-2\cr 0&1} \ .  \label{T1}
\end{equation}
Finally around $u=-1$ we have for the counter clockwise direction
\begin{eqnarray}
&&U_u(-1-r+i\epsilon)=\pmatrix{K(-\ft{r}2)+iK(1+\ft{r}2-i\epsilon)
\cr iK(1+\ft{r}2-i\epsilon) }=U_u(-1-r-i\epsilon)
+2\pmatrix{K(-\ft{r}2 )\cr iK(-\ft{r}2)}\nonumber\\
&& \Rightarrow \ T_{(-1)}= \pmatrix{3&-2\cr 2&-1} \ .    \label{Tm1}
\end{eqnarray}

We could have restricted ourselves to the computation of
the monodromy matrix in $u=1$, i.e. $T_{(1)}$, and $B$. Indeed,
since $B$ permutes the singular points $1,-1$ among themselves
the monodromy around any of them  can be obtained from the
monodromy at the  other point  by conjugation with  $B$.
Explicitly, from
considering  the paths in moduli space one has
\begin{equation}
\label{diedro_20}
T_{(-1)} = B T_{(1)} B=S_B T_{(1)} S_B^{-1}\ .
\end{equation}
Furthermore, the product of the monodromy matrices in all the
singular points must give the identity, as a contour encircling
all the singularities is homotopic to zero. Hence the monodromy
around the point $u=\infty$ is obtained from:
\begin{equation}
\label{diedro_21}
T_{(\infty)} T_{(1)} T_{(-1)}=\unity  \ .
\end{equation}

$T_\infty$ and $T_{(1)}$ generate the $\Gamma(2)\subset \Gamma=
PSL(2,\ZZ)$ group. This is the normal principal congruence subgroup
of order~2 of the modular group of the torus $\Gamma$:
\begin{equation}
\gamma \, \in \, \Gamma(2)~ :\quad \gamma
=\pmatrix{1+2 q & 2p\cr 2r &1+2 s} \ ;\qquad 2(qs-pr)+q+s=0 \ .
\end{equation}
It is known that $ D_3 \sim PSL(2,\ZZ_2 ) \sim \Gamma \, /
\Gamma(2)$.  So as the full $D_3$ is not realized as duality
symmetries, the final duality group is not $\Gamma$, but the subgroup
which can be generated by $S_B$ and $T_{(1)}$.
The perturbative part is the subgroup where the upper right component
is zero. It is generated by \eqn{pertmonoSW},
i.e. by  $T_{(\infty)}$ and $-S_B T_{(1)}^{-1}$.

\sectionsub{Supergravity}    \label{ss:sugra}
 Also for supergravity, I am first giving the original construction
 of the actions, before re-expressing the result in a symplectic
 invariant way. We will see that here the symplectic formulation is
 even more necessary than in the rigid case. Then I explain the
 connection to moduli of Calabi-Yau manifolds.

I treat here only the vector multiplets. For more complete reviews of
tensor calculus, the multiplets, and construction of the actions, see
\cite{KarpaczIITorino}. About Calabi-Yau manifolds, there is a fast
increasing literature. For a recent
introduction I refer to \cite{fresoriabook}.

\subsection{Vector multiplets coupled to supergravity }
\label{ss:vmcoupledSG}
To construct the action of vector multiplets coupled to supergravity,
I use the superconformal tensor calculus. This method keeps the
fruits of the superspace approach while avoiding a lot of technical
difficulties with many constraints on large superfields. No
superfields are introduced to covariantize the interactions of the
matter multiplets. This is replaced by defining the multiplets
within a larger algebra. The formulation can be compared with the
description of
non-abelian vector multiplets which I gave at the end of
section~\ref{ss:multiplets}. There, a full superspace approach was
replaced by defining the vector multiplet in the algebra
\eqn{naalgepsilon}. Still I keep working with multiplets, which shows
the structure of actions (and potentials) better due to the presence
of auxiliary fields. Of course, there is also a drawback. One is
never sure that the constructions one thinks of are the most general
one. In some cases the most general result is not found yet. As an
example I mention here that it remains a challenge to construct the
most general couplings of hypermultiplets in superspace or tensor
calculus.

The group which I am going to use for defining the multiplets is the
{\elevenit superconformal group}. This group is bigger than the
super-Poincar\'e group, which we want to have as final invariance of
the actions. The method constists of first constructing actions invariant
under the full superconformal group, and then afterwards choose
explicit gauge fixings. Fields which are just introduced to allow
later gauge choices are called compensating fields. The method
is called 'gauge equivalence' (see
the reviews \cite{revN2,KarpaczIITorino}). It has the advantage of
showing more
structure in the theory. In $d=4,\,N=2$ the superconformal group is
\begin{equation}
SU(2,2|N=2)\supset SU(2,2)\otimes U(1)\otimes SU(2)\ .\label{scgN2}
\end{equation}
The bosonic subgroup, which I exhibited, contains, apart from the
conformal group $SU(2,2)=SO(4,2)$, also $U(1)$ and $SU(2)$ factors.
The \Ka
ian nature of vector multiplet couplings and the \qu\ nature of
hypermultiplet couplings is directly related to the presence of
these two groups. The dilatations, special
conformal transformations, $U(1)\otimes SU(2)$, and an extra
$S$--supersymmetry in the fermionic sector will be broken by gauge
fixings. In that way we just keep the super-Poincar\'e invariance.

To describe theories as exhibited in table~\ref{tbl:multN2d4}, the
following multiplets are introduced: (other possibilities, leading
to equivalent physical theories, also exist, see
\cite{dWLVP,revN2,KarpaczIITorino}).
The {\elevenit Weyl multiplet} contains the
vierbein, the two gravitinos, and auxiliary fields. I introduce
{\elevenit $n+1$ vector multiplets} :
\begin{equation}
\left( X^I,\lambda^{iI},
{\cal A}_\mu^I  \right)\qquad\mbox{with}\qquad I=0,1,...,n.
\end{equation}
The extra vector multiplet labelled by $I=0$ contains the scalar
fields which are to be gauge--fixed in order to break dilations and
the $U(1)$, the fermion to break the $S$--supersymmetry, and the
vector which corresponds to the physical vector of the supergravity
multiplet in table~\ref{tbl:multN2d4}. Finally, there are {\elevenit
$s+1$ hypermultiplets}, one of these contains only auxiliary fields
and fields used for the gauge fixing of $SU(2)$. For most of this
paper I will not discuss hypermultiplets ($s=0$).

In the first step we have to build a superconformal invariant
action. To do so similarly to the construction in rigid superspace
\eqn{superspaceF}, the highest component of the chiral superfield
$F(\Phi)$ should have weight~4 under dilatations. This implies that
the lowest component $F(X)$ should have Weyl weight~2. The weight of
fields in a vector multiplet is fixed due to the presence of gauge
vectors which should have weight zero. This fixes the Weyl weight of
the scalars $X^I$ to~1. Combining the above facts leads to the
important conclusion that for the coupling of vector multiplets to
supergravity, one again starts from a holomorphic prepotential
$F(X)$, this time of $n+1$ complex fields, but now it must be a
{\elevenit homogeneous function of degree two} \cite{DWVP}.

In the resulting action appears
$-\ft12 i(\bar X^I F_I -X^I\bar F_I)eR$, where $R$ is the space--time
curvature. To have the canonical kinetic terms for the graviton, it
is therefore convenient to impose as gauge fixing for dilatations the
condition
\begin{equation}
i(\bar X^I F_I - \bar F_I X^I) = 1\,.\label{constraint}
\end{equation}
Therefore, the physical scalar fields parametrize an $n$-dimensional
complex hypersurface, defined by the condition \eqn{constraint},
while the overall phase of the $X^I$ is irrelevant in view of the
$U(1)$ in \eqn{scgN2} which we have not fixed.

 The embedding of this hypersurface can be described in terms of
$n$ complex coordinates $z^\alpha$ by letting $X^I$ be proportional to
some holomorphic sections $Z^I(z)$ of the projective space
$\Cbar \IP^n$ \cite{CdAF}.
The $n$-dimensional space parametrized by the
$z^\alpha $ ($\alpha =1,\ldots, n$) is a \Ka\ space; the K\"ahler metric
$g_{\alpha \bar \beta }=\partial_\alpha \partial_{\bar \beta }
K(z,\bar z)$ follows from the K\"ahler potential
\begin{eqnarray}
&&K(z,\bar z)=
-\log\Big[i \bar Z^I (\bar z)\,F_I (Z(z)) -i Z^I (z)\,
\bar F_I (\bar Z(\bar z))\Big] \ , \quad \mbox{ where }\label{KP}\\
&& X^I =e^{K/2}Z^I (z)\ ,\qquad\bar X^I =
e^{K/2}\bar Z^I (\bar z) \ .
 \nonumber
\end{eqnarray}
The resulting geometry is known as {\elevenit special} \Ka\ geometry
\cite{DWLPSVP,DWVP,special}.
\par
A convenient choice of inhomogeneous coordinates $z^\alpha $
are the {\elevenit special} coordinates, defined by
\begin{equation}
z^\alpha = \frac{X^\alpha }{X^0}\ ,\qquad \mbox{or}\qquad
Z^0(z)=1\,,\qquad Z^A (z) = z^\alpha \,.
\label{defspcoor}
\end{equation}
\par
In the general form of the spin-1 action \eqn{genL01}, the indices
$\Lambda$ are now replaced by $I$ running over $m=n+1$ values, as the
graviphoton is included. The matrix ${\cal N}$ is given by
\begin{equation}
\cN_{IJ}(z,\bar z) =\bar F_{IJ} +2i {{(\Im F_{IJ})( \Im
F_{JK})X^\G
X^K}\over{(\Im F_{LM})\ X^L X^M}}\ .
\label{Ndef}
\end{equation}
The kinetic energy terms of the scalars are positive definite in a
so-called 'positivity domain'. This is given by the conditions
$g_{\alpha {\bar\beta} }>0$ and $e^{-K}>0$. Then it follows that
$\Im{\cal N}_{IJ}<0$ \cite{BEC}. Note that the fact that the
positivity domain is non-empty, restricts the functions $F$ which can
be used. In \cite{trspring} some examples are discussed.
\subsection{Symplectic transformations}
To extend the duality transformations on the vectors to the scalar
sector, one introduces also here a $2m=2(n+1)$ component vector
\begin{equation}
V=\pmatrix{X^I \cr F_I} \ ,  \label{Vsugra}
\end{equation}
which transforms as a symplectic vector. Note that now the upper
part of this vector can not be used as independent coordinates, as
there are only $n$ scalars. However, in the formulation described in
the previous subsection the lower part still depends on the upper
components in the sense that $F_I(z, \bar z)=F_I(X(z, \bar z))$, as
$F_I=\frac{\partial F(X)}{\partial X^I}$. When symplectic
reparametrizations are performed, this can change. Indeed, now the
matrix $\frac{\partial \tilde X^I}{\partial X^J}$, see
\eqn{invertApBF}, may be non-invertible \cite{f0art} while the
inverse of $(A+B{\cal N})$ still exists because $\bar {\cal
N}_{IJ}\neq F_{IJ}$. This implies that the proof of existence of a
new function $\tilde F(\tilde X)$ is invalidated in these cases.
However, the transformed symplectic vector still describes the same
theory. It is clear that as the new $X^I$ are not independent, one
can not choose special coordinates \eqn{defspcoor} in terms of the
new symplectic vectors. So as in this new formulation the function
$F(X)$ does not exist, one can not immediately obtain this formulation
from superspace or tensor calculus. This shows that in supergravity,
even more
than in rigid symmetry, we need a formulation which does not start
from the function $F$, but is essentially symplectic covariant. This
will be treated in section~\ref{ss:defsgsg}.

In \cite{trspring} we gave a few simple examples of symplectic
reparametrisations and duality symmetries, including such a
transformation to a formulation without a function $F(X)$.
Another important example of this phenomenon is the description of
the manifold
\begin{equation}
 \frac{SU(1,1)}{U(1)}\otimes
\frac{SO(r,2)}{SO(r)\otimes SO(2)}\ .\label{manSTr}
\end{equation}
This is the only special \Ka\ manifold which is a product of two
factors \cite{sertoi}. Therefore it is also determined that in the
classical limit of the compactified heterotic string, where the
dilaton does not mix with the scalars of the other vector multiplets,
the target space should have that form.
The first formulation of these spaces used a function $F$ of the form
\cite{BEC}
\begin{equation}
F=\frac{d_{ABC}X^A X^B X^C}{X^0}  \ .
\end{equation}
In fact, such a form of $F$ is what one expects for all couplings
which can be obtained from $d=5$ supergravity \cite{GuSiTo}.
Such manifolds are
called `very special \Ka\ manifolds'. In such a formulation for
\eqn{manSTr} the $SO(r,2) $ part of the duality group sits not
completely in the perturbative part of the duality group, i.e. one
needs $B\neq 0$ in the duality group to get the full $SO(2,r)$.
However, from the superstring compactification one expects
$SO(2,r;\Zbar ) $ as a perturbative ($T$-duality) group.

By making a symplectic transformation this can indeed be obtained
\cite{f0art}. After that symplectic transformation one has a
symplectic vector $(X^I,F_I)$ satisfying
\begin{equation}
 X^I\,\eta_{IJ}\,X^J =0\ ;\qquad
F_I =S\,\eta_{IJ}\,X^J \ ,
\end{equation}
where $\eta_{IJ}$ is a metric for $SO(2,r)$. The first constraint
comes on top of the constraint \eqn{constraint}, and thus implies
that the variables $z$ can not be chosen between the $X^I$ only.
Indeed, $S$ occurs only in $F_I$.
\pagebreak[2]
\subsection{Symplectic definition of special geometry}
\label{ss:defsgsg}
\nopagebreak[4]
The symplectic formulation of special geometry
was first given in the context
of a treatment of the moduli space of Calabi-Yau three-folds
\cite{special,FerStroCand,CdAFTrieste}. The formulation which we
present is based on \cite{CDFLL}, and the equivalence with our
previous formulation was explained in detail in \cite{prtrquat}.

The definition can again be given on the basis of a matrix ${\cal
V}$. As I already said, the symplectic vectors now have $2(n+1)$
components. Similarly ${\cal V}$ will be a $2(n+1)$ by $2(n+1)$
matrix. It consists of the following rows of symplectic
vectors:
\begin{equation}
{\cal V} = \pmatrix{ \bar V^T\cr U^T_\alpha \cr V^T\cr
\bar U^{\alpha\,T}  \cr}\ ,
\end{equation}
which again should satisfy \eqn{cVsympl}. Furthermore there are the
differential equations \eqn{flatness}. In \cite{prtrquat} the
covariant derivatives in that equations included connection for \Ka\
transformations. Here, I include this connection in the matrix ${\cal A}$.
This leads to
\begin{eqnarray}
{\cal A}_\alpha  &=&
\pmatrix{\ft12\partial_\alpha K&0&  0&0\cr \noalign{\vskip1mm}
0&-\ft12\delta_\beta^\gamma\partial_\alpha K& 0&
C_{\alpha \beta \gamma }\cr
\noalign{\vskip1mm}
 0&\delta_\alpha ^\gamma &-\ft12\partial_\alpha K&0\cr
 \noalign{\vskip1mm}
 \delta^\beta _\alpha &0  &0&
 \ft12\delta^\beta_\gamma\partial_\alpha K\cr }\nonumber\\
{\cal A}_{\bar \alpha } &=&
\pmatrix{-\ft12\partial_{\bar \alpha} K&0&0&g_{\bar \alpha  \gamma }\cr
\noalign{\vskip1mm}
 0&\ft12\delta_\beta^\gamma\partial_{\bar \alpha} K&
 g_{\bar \alpha \beta }&0 \cr
\noalign{\vskip1mm}
  0&0  &\ft12\partial_{\bar \alpha} K&0\cr \noalign{\vskip1mm}
 0& \bar C_{\bar \alpha }^{\, \beta \gamma }  &0&
 -\ft12\delta^\beta_\gamma\partial_{\bar \alpha} K\cr } \,.
\end{eqnarray}
Here $K$ is the \Ka\ potential.
Again the system of equations is invariant under constant symplectic
transformations acting from the right on ${\cal V}$ and gauge
transformations of the form \eqn{transfS} on the left. The latter can
be used again to obtain holomorphic equations \cite{CDFLL}.  E.g.
with $S=\exp \left(\ft12 \mbox{diag}(-K,K,K,-K)\right) $ one removes
the \Ka\ connections from ${\cal A}_{\bar \alpha}$ and these
equations thus imply holomorphicity for the vector $V$ in that basis,
whose first components are then in fact the $Z^I$ from \eqn{KP}.

As in the rigid case, one can obtain equations similar to
\eqn{diffeqnC} (corrected by the \Ka\ connection) and the curvature:
\begin{equation}  R^\alpha {}_{\beta \gamma}{}^\delta
=2\delta^\alpha_{(\beta}\delta^\delta_{\gamma)}-
C_{\beta \gamma \epsilon}\bar C^{\alpha\delta\epsilon}  \ ;\qquad
\bar C^{\alpha\delta\epsilon}
\partial_{\bar \alpha}\left( e^{-K}C_{\alpha\beta\gamma}\right) =0\ ;
\qquad {\cal D}_{[\alpha}e^KC_{\beta]\gamma\delta}=0\ .
\label{Rlocal}\end{equation}

\subsection{Calabi-Yau manifolds}
Calabi-Yau manifolds are
complex manifolds with Ricci tensor which is exact, i.e.
$R_{\alpha\bar\beta}=\partial_\alpha A_{\bar\beta} -
\partial_{\bar\beta} A_\alpha$.    For our purposes we need
3-folds (i.e. real dimensional 6, related to the reduction of 6
dimensions of the superstring). The Hodge diamond gives the number of
closed and non-exact forms of all holomorphic types. E.g. for the
Riemann surface of genus $n$ it looks like
\begin{equation}
\begin{array}{ccc} &h^{00}&\\ h^{10}&&h^{01} \\ &h^{11}& \end{array}
\qquad =\qquad
\begin{array}{ccc} &1&\\ n&&n \\  &1& \end{array}
\end{equation}
For the Calabi-Yau
manifolds this Hodge diamond is already fixed for a large part:
\begin{equation}
\begin{array}{ccccccc}
 & &  & 1      &   &  & \\
 & &0 &        & 0 &  & \\
 &0&  &h^{11}&   & 0& \\
1& &h^{12}&  &h^{21}&&1\\
 &0&  &h^{22}&   & 0& \\
 & &0 &        & 0 &  & \\
 & &  & 1      &   &  &
\end{array}
\ ;\qquad\mbox{with}\qquad \begin{array}{l}
h^{11}=h^{22}\\
h^{12}=h^{21}\ .
\end{array}
\end{equation}
A large class of them can be obtained as the vanishing locus of a
quasi-homogeneous polynomial in projective space \cite{Candelaspshom},
similar to the
description of Riemann surfaces in section~\ref{ss:moduliRS}. For the
connection with special geometry one now considers all 3-forms.
Identifying $h^{12}=h^{21}=n$, there are $2(n+1)$ of these. The
integrals over a canonical basis of 3-cycles defines then  the
period matrix. One can obtain again
differential equations (Picard-Fuchs equations)  by
differentiating with respect to the moduli of the surface, which
are to be compared with the defining equations of special geometry as
given above.
As in the case of Riemann surfaces,
the duality symmetries are defined by the monodromies around
singular points and symmetries of the defining equation. For the
latter we do not have to restrict now to those having constant
rescaling factors \cite{monodrcy}.

\subsection{Special quaternionic manifolds and homogeneous special
manifolds} \label{ss:sperqu}

The \cmap\ \cite{CecFerGir} gives a mapping from a
special \Ka\ to a \qu\ manifold. It is induced by reducing
an $N=2$ supergravity action in $d=4$ space-time
dimensions to an action in $d=3$ space-time dimensions, by
suppressing the dependence on one of the (spatial) coordinates.
The resulting $d=3$ supergravity theory can be written in terms
of $d=3$ fields and this rearranges the original fields such that
the number of scalar fields increases from $2n$ to $4(n+1)$.  This
map is also obtained in string theory context by changing from a type
IIA to a type IIB string or vice-versa.

This leads to the notion of `{\elevenit
special quaternionic manifolds}', which are those manifolds appearing
in the image of the \cmap. Similarly,  very special
real manifolds are the manifolds defined by coupling (real) scalars
to vector multiplets in 5 dimensions \cite{GuSiTo}
(characterised by a symmetric tensor
$d_{ABC}$). Very special \Ka\ manifolds \cite{brokensi}
are induced as the image under the
\rmap\  (dimensional reduction from 5 to 4 dimensions)
and very special quaternionic manifolds as the image of
the \crmap.

It turns out that these very special manifolds
contain all known homogeneous non--symmetric \Ka\ and
quaternionic spaces. The classification of the homogeneous spaces is
related to the enumeration of all realizations of real Clifford
algebras, see \cite{ssss}, or the summary in \cite{prtrquat}. In
these papers also all the continuous isometries of homogenous special
and of very special manifolds are given.

\section{Acknowledgements}
I thank M. Bill\'o, A. Ceresole, R. D'Auria, B. de Wit, S. Ferrara,
P. Fr\`e, I. Pesando, T. Regge, P. Soriani and W. Troost for
clarifying discussions.
This work was carried out in the framework of the
project "Gauge theories, applied supersymmetry and quantum
gravity", contract SC1-CT92-0789 of the European Economic
Community.

\appendix
\renewcommand{\section}[1]{\refstepcounter{section}
\vglue 0.5cm
{\elevenbf\noindent Appendix \thesection . #1}
\vglue 0.4cm
\setcounter{subsection}{0}
}
\section{Conventions}

I use the metric signature $(-+++)$.
The curved indices are  denoted by
$\mu, \nu, \ldots = 0,\ldots , 3$ and the flat ones by $a, b, ...$.
(Anti)symmetrization is done with weight one: $A_{[ab]}=\ft12 \left(
A_{ab}-A_{ba}\right) $ and $A_{(ab)}=\ft12 \left(
A_{ab}+A_{ba}\right) $.

I define \begin{equation}
\epsilon^{\mu\nu\rho\sigma}=\sqrt{-g}\,e^\mu_a\,
e^\nu_b\, e^\rho_c \,e^\sigma_d \,\epsilon^{abcd} \ ; \qquad
\epsilon^{0123}=i\ ,
\end{equation}
where the former implies that the latter is true for flat as well as
for curved indices.
 I introduce the self--dual and anti--self dual tensors
\begin{equation}
F_{ab}^\pm =\ft12\left(F_{ab}\pm {}^\star F_{ab}\right)\qquad
\mbox{with}\qquad {} ^{\star} F^{ab}=\ft12 \epsilon ^{abcd}F_{cd} \ .
\end{equation}

The gamma and sigma matrices are defined by
\begin{equation}
\gamma_a \gamma_b=\eta _{ab}+2\sigma_{ab} \ , \qquad
\gamma_5=i \gamma_0\gamma_1\gamma_2\gamma_3\ ,  \label{defgamma}
\end{equation}
which implies that $\ft12\epsilon^{abcd}\sigma_{cd}=-\gamma^5
\sigma^{ab}$.
The following realization brings you to the 2-component formalism:
(here $\alpha=1,2,3$, but this is not used anywhere else)
\begin{equation}
\gamma_0=\pmatrix{0&i\unity _2\cr i\unity _2 & 0\cr}\ ;\qquad
\gamma_\alpha=\pmatrix{0&-i\sigma_\alpha\cr i\sigma_\alpha & 0\cr}\ ; \qquad
\gamma_5=\pmatrix{1&0\cr 0& -1 \cr}\ .\label{2comprep}
\end{equation}
The matrices $\gamma_\alpha$ and $\gamma_5$ are hermitian, while
$\gamma_0$ is antihermitian.

There is a charge conjugation matrix $\cal C$ such that
\begin{equation}
{\cal C}^T=-{\cal C} \ ; \qquad {\cal C}\gamma_a{\cal C}^{-1}=
-\gamma_a^T\ .
\end{equation}
This fixes ${\cal C}$ up to a (complex) constant.
One can fix the proportionality constant (up to a phase) by demanding
${\cal C}$ to be unitary, so that ${\cal C}^*=-{\cal C}^{-1}$.
In the representation \eqn{2comprep} we can choose
\begin{equation}
{\cal C}=\pmatrix{\epsilon_{AB} & 0 \cr
0& \epsilon^{\dot A \dot B} \cr}\ ,
\end{equation}
where $\epsilon$ is the antisymmetric symbol with $\epsilon_{AB}=
-\epsilon^{\dot A \dot B}=1$ for $A=1,\,B=2$.

Majorana spinors are spinors $\chi $ which satisfy the
`reality condition' which says that their `Majorana conjugate' is
equal to the `Dirac conjugate'
\begin{equation}
\bar \chi \equiv \chi ^T{\cal C}=-i\chi^\dagger \gamma _0 \equiv
\bar \chi^C \qquad
\mbox{ or }\qquad \chi^C\equiv -i\gamma_0 {\cal C}^{-1}\chi^*=\chi\ .
\label{Majorana}
\end{equation}
The factor $-i$ is just a conventional choice as the
phase of ${\cal C}$ is arbitrary.
Majorana spinors can thus be thought as spinors $\chi_1 + i \chi_2$,
where $\chi_1$ and $\chi_2$ have real components, but these are
related by the above condition. There exists a Majorana representation
where the matrices $\gamma_\mu$ are real, and with a convenient
choice of the phase factor of ${\cal C}$ the Majorana spinors are
just real.

In the 2-component formulation, indices are raised or lowered with
$\epsilon$ in a NW-SE convention
\begin{equation}
\chi^A=\epsilon^{AB}\chi_B\ ;\qquad \chi_A= \chi^B\epsilon_{BA}\ ,
\end{equation}
which implies $\epsilon^{AB}\epsilon_{BC}=-\delta^A{}_C$, and
the Majorana condition for a spinor $(\zeta^A, \zeta_{\dot
A})$  is then $\zeta_A=\left(  \zeta_{\dot A}\right)^*$.

I often use Weyl spinors, where the left and right chiral spinors
are defined as
\begin{equation}
\chi_L =\ft12 (1+\gamma_5) \chi\ ; \qquad
\chi_R =\ft12 (1-\gamma_5) \chi\ .
\end{equation}
In extended supergravity the chirality is
indicated by the position of the $i,j$ index
(index running over $1,\ldots N$ for $N$-extended
supergravity). To choice of chirality for the spinor with an upper
(lower) index can change for each spinor. It is chosen conveniently
on the first occurrence of the spinor.
E.g. in this paper:
\begin{equation}\begin{array}{llll}
\epsilon^i=\gamma_5 \epsilon^i& Q^i=-\gamma_5 Q^i&
\theta^i=\gamma_5\theta^i&\\
\Psi^i=-\gamma_5 \Psi^i& \Lambda^i=-\gamma_5 \Lambda^i&
 \Omega^i=-\gamma_5 \Omega^i & \phi^i=\gamma_5 \phi^i
\end{array}
\end{equation}
Note that these Weyl spinors are not Majorana.
A spinor which is Majorana and Weyl with the above definitions would
only be possible in $d=2$ mod 8. In the two-component representation,
the left Weyl spinors are just $(\zeta^A,0)$.

Hermitian conjugation on a bispinor
reverses by definition the order of the spinors.
To perform h.c. in practice it is easier to replace it by charge
conjugation. For any bispinor one has
\begin{equation}
( \bar \chi M \lambda)^\dagger = (\bar \chi^C M^C\lambda^C)\ ,
\end{equation}
where $C$ was defined for the spinors in \eqn{Majorana}, and for the
matrix $M$ one has $M^C=-\gamma_0 {\cal C}^{-1}M^*{\cal C}\gamma_0$.
The gamma matrices are inert under this transformation, but
$\gamma_5$ changes sign due to the $i$ in its definition \eqn{defgamma}.
The Majorana spinors are inert under $C$, but the Weyl spinors change
chirality: $\epsilon_i^C=\epsilon^i$. Therefore h.c. effectively
replaces $i$ by $-i$, interchanges upper and lower indices, and
$\epsilon^{\mu\nu\rho\sigma}$ changes sign, or self--dual becomes
anti--self dual.

In exceptional cases I use spinor indices. Then by definition $\bar
\chi \lambda= \chi^\alpha  \lambda_\alpha$, where $\chi^\alpha=
\chi_\beta {\cal C}^{\beta\alpha}$ is now the counterpart of
\eqn{Majorana}. Its inverse is ${\cal C}^{-1}_{\alpha\beta}$.
The $\gamma$--matrices are written as $\left(
\gamma_\mu\right) _\alpha{}^\beta$.
%
%
\par\pagebreak[3]
\noindent{\bf \large Exercises or useful formulae}.\par
\nopagebreak[4]
Here all spinors are fermionic and the left or right projections
of Majorana spinors.
\begin{equation}\begin{array}{lll}
\sigma^{ab}F_{ab}\epsilon^i=\sigma^{ab}F_{ab}^-\epsilon^i
& F^+_{\mu\nu}G^{-\mu\nu}=0
& \left( F^+_{\mu\nu}G^{+\mu\nu}\right) ^\dagger =
F^-_{\mu\nu}G^{-\mu\nu}      \\
\bar  \epsilon^i\equiv \bar \epsilon^i\gamma_5
&\left( \bar \xi  _L \lambda_L \right) ^\dagger =
\bar \xi  _R\lambda_R&\\
 \bar \xi  _L \lambda_L= \bar \lambda_L \xi  _L
& \bar \xi  _L \sigma_{ab}\lambda_L=
- \bar \lambda_L \sigma_{ab}\xi  _L
& \bar \xi  _L \gamma_a\lambda_R= -\bar \lambda_R \gamma_a
\xi  _L \\
 \lambda_L \bar \chi_ R=-\ft12 \gamma^m  (\bar \chi_R \gamma_m
\lambda_L )
& \lambda_L \bar \chi_L = -\ft12 \unity (\bar \chi_L \lambda_L )+
\ft12 \sigma^{ab} (\bar \chi_L\sigma_{ab} \lambda_L ) &
\end{array}  \label{antisdsigma}
\end{equation}
Show that in $N=2$ in a commutator
$\left[\delta(\epsilon_1),\delta(\epsilon_2)\right]$ only the
following index structures and their hermitian conjugates can appear:
\begin{equation}
\bar \epsilon_1  ^{[i}\epsilon_2^{j]}\ ;\qquad
\bar \epsilon_1  ^{i}\gamma_a\epsilon_{2j}\ ;\qquad
\bar \epsilon_1  ^{(i}\sigma_{ab}\epsilon_2^{j)}\ .
\end{equation}

\section{Normalisations}

Unfortunately the normalization of $F$ and various other
functions vary in the $N=2$ literature. In
table~\ref{tbl:compnot}, I compare the notations of various articles
(in supergravity).
The first column is the notation
used here, in \cite{trspring,prtrquat}, and most of it also in
\cite{dWKLL}. The column 'old' refers to the articles
\cite{DWLPSVP,DWVP,structure,KarpaczI,dWLVP,brokensi,ssss,CecFerGir,%
KarpaczIITorino,revN2,BEC}.
Note that the first row shows that also the convention of the
space-time metric has changed.  The freedom of the real parameter $\alpha$,
indicated in the second column, can be repeated in all columns, but
looks most useful in this case.

\begin{table}[ht]\caption{Comparison of notations}
\label{tbl:compnot}\begin{center}\begin{tabular}{|ccccc|}\hline
Here& old  & \cite{CdAF} & \cite{f0art,CdAFTrieste}&\cite{fresoriabook} \\
\hline
 $g_{\mu\nu}$& $g_{\mu\nu}$&$-g_{\mu\nu}$&$-g_{\mu\nu}$&$g_{\mu\nu}$\\
 $g_{\alpha\bar \beta}$&$g_{A\bar B}$&$g_{ij^*}$&$g_{i\jb}$&$g_{ij^*}$ \\
$K$&        $ -K+2\log \alpha $ &$G$& $K$ & ${\cal K}$\\
$F$ & $-\ft i4 F$ & $-iF$& $F$ & $-F$ \\[1mm]
$X^I$ & $\alpha X^I$ & $L^\Lambda$ & $L^\Lambda$   & $L^\Lambda$  \\
$Z^I$ &   $Z^I$ &   $X^\Lambda$  &   $X^\Lambda$   & $X^\Lambda$\\
$F_I$ & $-\ft i{4\alpha}F_I$&$-iF_\Lambda$ & $M_\Lambda$ &
$-F_\Lambda$ \\[2mm]
$\Im F_{IJ}$ & $-\ft 1{2\alpha^2}N_{IJ}$&$-\ft12 N_{\lambda\sigma}$&
$\Im F_{\Lambda\Sigma}$ & $\ft i2 N_{\Lambda\Sigma}$\\[2mm]
${\cal N}_{IJ}$&$\ft i{\alpha^2}{\cal N}_{IJ}$&$
-i{\cal N}_{\Lambda\Sigma}$&${\cal N}_{\Lambda\Sigma}$&\\[2mm]
$C_{\alpha\beta\gamma}$& $e^{-K}Q_{ABC}$&$C_{ijk}$&$i C_{ijk}$&$-i
C_{ijk}$ \\[1mm]
${\cal F}_{\mu\nu}^I$&$\alpha F_{\mu\nu}^I$&$ 4F_{\mu\nu}^\Lambda$&
$\sqrt{2}{\cal F}_{\mu\nu}^\Lambda$  & \\[1mm] \hline
\end{tabular}\end{center}\end{table}

The symplectic matrices compare as follows between the notations here
(left hand side) and in the 'old' notation (right hand side):
\begin{equation}
\pmatrix{A&B\cr C&D\cr}=
\pmatrix{U& 2\alpha^2 Z\cr \ft1{2\alpha^2}W & V\cr}
\end{equation}

\section{The volume form}\label{app:volform}
I consider here a pseudo homogeneous space in $N+2$ dimensions with
coordinates $X^\Lambda$, where the weight of $X^\Lambda$ is
$\beta_\Lambda$. I will prove that the volume form \cite{Morrison}
\begin{equation}
\omega =X'^\Omega \, \dXN0\, \epsilon_{\Omega\Sigma_0\cdots \Sigma_N} \ ,
\label{defomega}\end{equation}
where I introduced $X'^\Omega= \beta_\Omega X^\Omega$, satisfies
\eqn{propomega} when the overall degree of its left hand side is
zero.
The degree of $\omega$ is $\gamma\equiv \sum_{\Lambda}
\beta_\Lambda$, and therefore the degree of ${\cal P}^\Lambda$ is
$\alpha_\Lambda= \beta_\Lambda -\gamma$.
This implies, using the notation with primes,
\begin{equation}
X'^\Omega\partial_\Omega {\cal P}^\Lambda(X)= \alpha_\Lambda {\cal P}^\Lambda
\label{homogeneityY}\end{equation}
(no sum over $\Lambda$ in r.h.s.).
Let me take
\begin{equation}
\Theta= -(N+1)\,{\cal P}^\Lambda X'^\Omega\,\dXN1 \,
\epsilon_{\Lambda\Omega\Sigma_1\cdots \Sigma_N}    \ .
\end{equation}
The differential gives
\begin{eqnarray}
\dop\Theta&=&- (N+1)\left( \partial_{\Sigma^0}{\cal P}^\Lambda\right) \,
X'^\Omega\,\dXN0 \,
\epsilon_{\Lambda\Omega\Sigma_1\cdots \Sigma_N}\nonumber\\
&& -(N+1){\cal P}^\Lambda \,\dop  X'^{\Sigma^0}\wedge \dXN1  \,
\epsilon_{\Lambda\Sigma_0\cdots \Sigma_N} \ .
\end{eqnarray}
The last term involves
\begin{equation}
\dop  X'^{\Sigma^0}\wedge \dXN1=\beta_{\Sigma^0} \dXN0
\end{equation}
By symmetrization, the $\beta_{\Sigma^0}$ becomes replaced by the
average $\beta$ of those in the differentials, which is all but
$\Lambda$. Thus, taking into account the Levi-Civita tensor, it can
be replaced by $\frac{\gamma-\beta_\Lambda}{N+1}=-\frac{\alpha_\Lambda}
{N+1}$. On the first term of $\dop\Theta$, I use the Schouten
identity in the $N+3$ lower indices. This gives
\begin{eqnarray}
\dop\Theta&=&  \left( \partial_\Lambda {\cal P}^\Lambda\right) \omega
\nonumber\\ &&
+\left( - X'^\Omega \partial_ \Omega {\cal P}^\Lambda
+\alpha_\Lambda {\cal P}^\Lambda
\right) \dXN0 \epsilon_{\Lambda\Sigma_0\cdots \Sigma_N}\ ,
\end{eqnarray}
which, combined with \eqn{homogeneityY}, gives the desired result.
\pagebreak[4]
\section{Some formulas about elliptic integrals.}  \label{app:ellint}
First a remark: I use the notation $K(x)$ for what is usually denoted
as $K(k)$ where $x=k^2$, and similar for other elliptic integrals.
The elliptic integrals are in general defined as
\begin{equation}
J(x)=\int_{0}^{\pi \over 2}  \frac{\Phi}{\sqrt{1-x\sin ^2\theta}}\,
d \theta =\int_0^1 \frac{\Phi}{\sqrt{1-y^2}\sqrt{1-xy^2}}dy
\end{equation}
where $\Phi$ takes different values, see below, the second line is
obtained from $y=\sin \theta$,  and the square root
is positive for $0<x<1$.
For larger values of $x$ on the real axis, $y$ goes through the pole
at $y=1/k$, and we will have to choose in which way to encircle this
point, i.e. whether $x$ is just below or above the real axis.
So, the elliptic integrals have a branch cut going from $x=1$ to
$x=+\infty$ along the real line.

The values of $\Phi$ are
\begin{eqnarray}
\Phi=1 && J(x)\rightarrow K(x)=
{\pi \over 2} \, F({1\over 2},{1\over 2},1;{x})\nonumber\\
\Phi=1-x\sin^2\theta=1-xy^2 &&J(x)\rightarrow E(x)=
{\pi \over 2} \, F({1\over 2},-{1\over 2},1;{x})\nonumber\\
\Phi=\cos^2\theta=1-y^2 && J(x)\rightarrow  B(x)=
{\pi \over 4} \, F({1\over 2},{1\over 2},2;{x})
\end{eqnarray}
The relation between the functions $\Phi$ implies
\begin{equation}
x\,B(x)= E(x) -(1-x) K(x)\ .
\end{equation}
The integral of $K(x)$ is
\begin{eqnarray}
\ft12 \int_0^x K(t)\,dt&=& \int_0^{\pi/2}\frac{d\theta}{-\sin^2\theta}
\left( \sqrt{1-x\sin ^2\theta} -1\right) \nonumber\\
&=&\left[\frac{\cos \theta}{\sin\theta}
\left( \sqrt{1-x\sin ^2\theta} -1\right)\right]_0^{\pi/2}  -
\int_0^{\pi/2}\frac{\cos \theta}{\sin\theta} d \sqrt{1-x\sin ^2\theta}
\nonumber\\
&=& \int_0^{\pi/2}  \frac{x\cos^2\theta}{\sqrt{1-x\sin ^2\theta}}\,
d \theta =x B(x)\ . \label{intK}
\end{eqnarray}

To obtain $K$ for arguments near the real line higher than 1, consider
(I still use $0<x<1$, and define $k=\sqrt{x}>0$)
\begin{eqnarray}
\frac{1}{k}K(\frac{1}{x}\pm i\epsilon)&=&\int_0^1 \frac{\frac{1}{k}dz}
{\sqrt{1-z^2} \sqrt{1-\frac{1}{x\mp i\epsilon}z^2}}
\nonumber\\
&=&
\int_0^k \frac{\frac{1}{k}dz}
{\sqrt{1-z^2} \sqrt{1-\frac{1}{x}z^2}} +
\int_k^1 \frac{\frac{1}{k}dz}
{\sqrt{1-z^2}(\mp i)\sqrt{\frac{1}{x}z^2-1}}\ .
\end{eqnarray} In the last line, I
have split the integral in the region $z<k$ and $z>k$. For the latter, I
have gone around the pole at $z=k\mp i\epsilon$, which produces the phase
$e^{\mp i\pi/2}$ for the square root.
For the first term I now use the substitution $y=z/k$,
while for the second term I use $ y^2=\frac{1-z^2}{1-x}$. This leads
to
\begin{equation}
K(\ft{1}{x}\pm i\epsilon) =k\left(K(x)\pm iK(1-x)\right) \ .
\label{3ptK}
\end{equation}
Following the same steps for $E(\ft1x)$ gives
\begin{eqnarray}
k E(\ft1x\pm i\epsilon)&=&xB(x)\mp i(1-x)B(1-x) =
\ft12\left[\int_0^x K(t)dt\pm i\int_1^x
K(1-t) dt  \right]\nonumber\\
&=&\ft12 \int_0^x \ft1{\sqrt{t}} K(\ft1t\pm i\epsilon) dt\ \mp \ iB(1)
\qquad\mbox{and}\qquad B(1)=1 \ .
\label{intE1x}
\end{eqnarray}

\eqn{3ptK} relates $K$ in 3 points. Using the same relation with $x$
substituted by one of the other values, relates $K(x)$ also to its
value in the points
\begin{equation}
x, \ 1-x , \ \frac{1}{x}, \ \frac{1}{1-x}, \ 1-\frac{1}{x}, \
-\frac{x}{1-x}\ . \label{6ptsK}
\end{equation}
This relation between 6 points shows the $D_3$ symmetry.
For $0<x<1$, the first two values are between 0 and 1, the next two
are higher than 1, and the last two are negative. However, in moving
$x$ to the other values,
one has to be careful
that none of the 3 points $x,\, 1-x$ and $1/x$ goes through the branch
cuts,
which would move it to another sheet.
There is the branch cut of the elliptic integral, but also the branch
cut of
$k=\sqrt{x}$, which goes from 0 to $-\infty$ along the real axis
in this equation. Considering this, one can see that for using \eqn{3ptK}
with the upper sign the point $x$ should be with negative
imaginary
part. Otherwise the 3 points are not on the first Riemann sheet.
For the lower signs the opposite domain should be used. With
$0<x<1$, this matters only for the points $\ft1x$ and $\ft1{1-x}$,
and for the analytic continuation of $k=\sqrt{x}$. E.g. for
$x\rightarrow 1-\frac{1}{x}$ this implies that one has to replace
$\sqrt{k}$ by
$\mp i \sqrt{\frac{1}{x}-1}$. Finally it turns out that the
independent relations are \eqn{3ptK}, the same with $x$ replaced
by $1-x$, and the relations
\begin{equation}
K(1- \frac{1}{x}  )=\sqrt{x}\,K(1-x)\ ;\qquad
K(-\frac{x}{1-x})=\sqrt{1-x}\,K(x)\ . \label{relKextra}
\end{equation}

Concerning the monodromies, the function $K$ has only a branch
point in $x=1$. From \eqn{3ptK} (using also \eqn{relKextra}) we have
\begin{equation}
K(\ft{1}{x}-i\epsilon)= K(\ft{1}{x}+i\epsilon) -2\,i\,K(1-\ft1x)\ .
\end{equation}
Writing $y\equiv \ft1x+i\epsilon=1+r$, we have
$y'\equiv 1+re^{2\pi i}= \ft1x-i\epsilon$, and
\begin{equation}
K(y')=K(y)-2\,i\,K(1-y)\ . \label{monoK}
\end{equation}
\vglue 0.5cm
\pagebreak[2]
{\elevenbf\noindent References}
\vglue 0.4cm

\end{document}